# Computational Insights into PEMFC Durability: Degradation Mechanisms, Interfacial Chemistry, and the Emerging Role of Machine Learning Potentials

*Jack Jon Hinsch\*[1], Kazushi Fujimoto[1,2]*

1) Organization for Research and Development of Innovative Science and Technology, Kansai University, 3-3-35 Yamate-cho, Suita, Osaka 564-8680, Japan

2) Department of Chemistry and Materials Engineering, Faculty of Chemistry, Materials and Bioengineering, Kansai University, 3-3-35 Yamate-cho, Suita, Osaka 564-8680, Japan.

email: r796859@kansai-u.ac.jp

Funding: The authors would like to thank the JSPS organisation for their support with the grant "Japan Society for the Promotion of Science" (Number: 25KF0283), and Fellowship award (Number: P25711).

Acronyms:

Proton exchange membrane fuel cell (PEMFC)
Catalyst layer (CL)
Gas diffusion layer (GDL)
Density functional theory (DFT)
Molecular dynamics (MD)
Machine learning (ML)
Ab initio molecular dynamics (AIMD)
General gradient approximations (GGA)
Standard hydrogen electrode (SHE)).
Beginning of life (BOL)
Accelerated stress test (AST)
End of life (EOL)
Carbon oxidation reaction (COR)
Deep potential molecular dynamics (DPMD)
Oxygen reduction reaction (ORR)
Electric double layer (EDL)
Hydrogen oxidation reaction (HOR)
Coordination number (CN)
Perfluoro sulfonic acid (PFSA)

# Abstract

Proton exchange membrane fuel cells (PEMFCs) are a promising clean energy technology, offering high efficiency and near-zero operational emissions for stationery and automotive applications. However, their widespread adoption remains limited by insufficient durability, driven by the degradation of the catalyst layer and proton exchange membrane under realistic operating conditions. While the macroscopic consequences of degradation are well established experimentally, the atomistic and molecular mechanisms that initiate and propagate failure remain incompletely understood. This review synthesizes recent advances in computational modelling, spanning density functional theory, molecular dynamics, and emerging machine learning potentials, to examine how chemical, mechanical, electrochemical, and contamination driven degradation mechanisms operate across multiple length and time scales. Key topics include radical-induced membrane degradation, platinum dissolution and carbon support corrosion, mechanical fatigue under electrical and hygrothermal cycling, and the impact of ionic and gaseous contaminants. A central finding is that these degradation pathways are not independent, but form strongly coupled feedback loops that no existing computational framework has been designed to capture this coupling simultaneously. Future directions are proposed, with emphasis on multiscale modelling frameworks and the application of machine learning interatomic potentials to the electrified interface.

# 1. Introduction

The growing global energy demand and the environmental harm caused by fossil fuels have accelerated the development of sustainable and eco-friendly energy solutions. Among various technologies, the proton exchange membrane fuel cell (PEMFCs) have gained significant attention due to their high efficiency, low operating temperature, environmental impact and dual applications for hydrogen production and electricity production.[1] PEMFCs are highly efficient and compact technologies that provide fast response times, leading to many potential applications.[2] Their stacking efficiency is also high, as one additional fuel cell in sequence increases its generated volumetric power density to 4.5 kW L$^{-1}$.[3]

Within the device, electricity originates from the conversion of hydrogen gas into water.[4,5] In the anode, hydrogen gas is collected and passes through an electrolyte membrane as protons, while the electrons pass through external circuitry to power devices (see **Figure 1**). Within the cathode, water generation occurs at a relatively slow rate, which defines the PEMFC's performance. The high energy density of the reaction attracts many investors, but it also presents specific technical challenges. The durability and reliability under dynamic operating conditions severely limit the technology's commerciality. There are several forms of this degradation, including the catalysis loss, membrane degradation and delamination of the membrane and catalyst layer (CL). Voltage cycling and start-stop cycles are a major accelerator of performance loss and deterioration.[3,6] Dynamics load profiles further increase the risk of catalyst loss and membrane degradation.

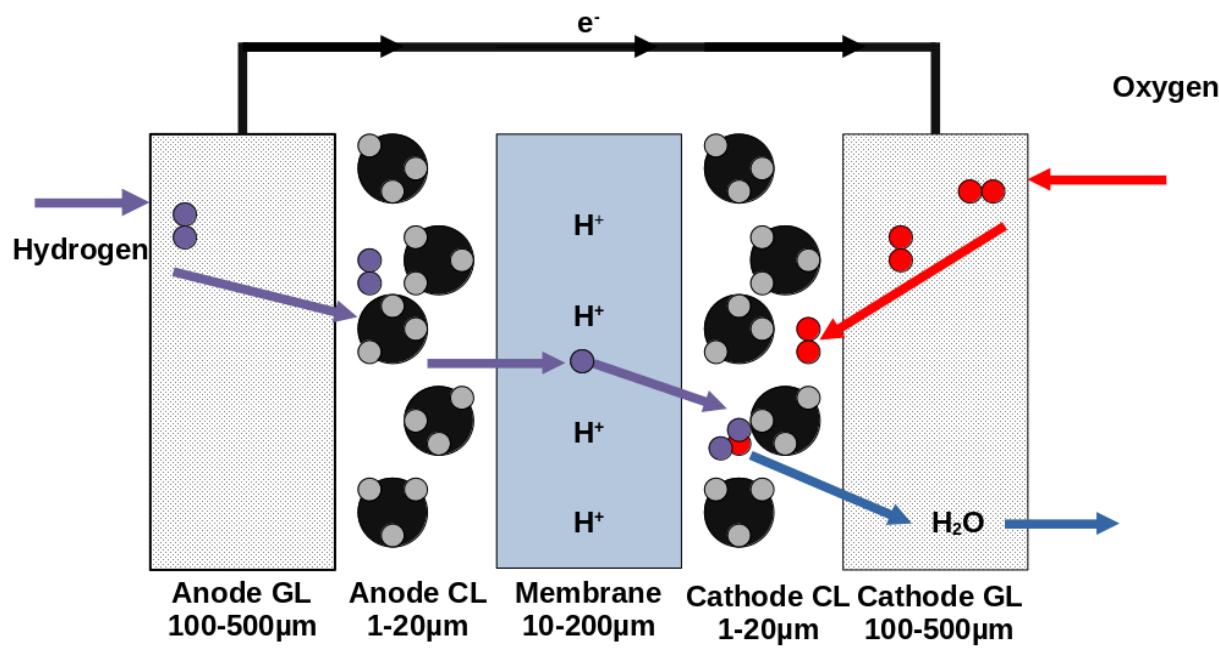

**Figure 1**: Visualization of the PEMFC gas diffusion layer (GDL) where the electricity producing reaction occurs. Grey circles are platinum, black circles are carbon supporting the platinum catalyst, red is oxygen, and purple is hydrogen. GL is the gas layer, and CL is catalyst layer. The membrane mainly contains the polymer and hydrogen protons that travel through to meet with oxygen atoms after splitting in the anodic CL. Once the hydrogen protons reach the cathodic CL they react with the a single O atom to form water. The electricity is produced from the lost electron during $H = e^- + H^+$ recombing with water during $2H + O_2 = 2H_2O$.

While the macroscopic consequences of PEMFC degradation are well established through decades of experimental investigation,[7–9] a mechanistic understanding at the atomistic and molecular scale has proven far more elusive. Identifying which bonds break first, how radical species propagate through the polymer backbone, and why the catalyst layer fails preferentially at certain interfaces requires tools that can resolve chemistry at the ångström and femtosecond scale a resolution that experiment alone cannot provide. Only recently have computational methods advanced sufficiently to address these questions directly. The rapid development of density functional theory (DFT), molecular dynamics (MD), and machine learning (ML) based approaches has begun to illuminate degradation mechanisms that were previously inaccessible, from the energetics of platinum dissolution and radical induced side-chain scission to the structural evolution of the Nafion-catalyst interface under mechanical and electrochemical stress.

However, a critical limitation has emerged from this body of work: the dominant computational strategy has been to study each degradation mechanism in isolation, under idealised conditions that do not reflect the complexity of an operating fuel cell. In reality, chemical, mechanical, and electrochemical degradation do not proceed independently. Instead, they form strongly coupled feedback loops in which each failure mode accelerates the others.[10] No existing computational framework captures this coupling simultaneously, and this gap between isolated mechanism studies and the coupled reality of degradation represents the defining unresolved challenge for the field.

This review, therefore, takes a critical and synthetic view of recent computational advances in PEMFC degradation modelling, to identify not only what has been learned but where the boundaries of current methods lie. Continuing from here, Section 1.1 introduces the computational methods employed across the field and establishes the length and time scales they address. Sections 2 and 3 examine chemical degradation of the membrane and catalyst layer, respectively, tracing radical attack pathways, platinum dissolution mechanisms, and carbon support corrosion. Section 4 addresses the effect of contaminants on the fuel cell performance. Section 5 covers the mechanical degradation of the CL and membrane, focusing on temperature and humidity. Section 6 examines the interconnected properties and synergistic impacts of the membrane and CL. Finally, section 7 describes alternatives and emerging directions for Nafion polymers.

## 1.1. The Role of Computational Methods in Understanding PEMFC Degradation

PEMFC degradation encompasses a range of chemical, mechanical, and electrochemical processes that collectively reduce fuel cell performance and longevity. The economic viability of PEMFCs for stationary applications demands that the technology lasts 5 years or over 40,000 hours of continuous operation.[11] In contrast, the automotive industry requires that fuel cell technology focus on low cost, practical lifetimes of 5,000 hours.[12] However, the automotive industry faces unique challenges, including startup/shutdown cycles, air intake in the fuel supply, and high operating temperatures.[13]

While each degradation mode has distinct origins, they are strongly interlinked: chemical degradation of Nafion leads to loss of mechanical strength; catalyst dissolution alters local potentials and promotes further radical formation; and interfacial delamination impairs electron and proton transport. The two critical components of the fuel cell are often the first targeted by these stresses. The platinum/carbon catalyst layer is the most expensive component of the fuel cell, and there are few acceptable alternatives. Membrane failure is a significant challenge for both stationary and mobile fuel cells. Understanding these degradation processes is crucial to the development of longer lasting membranes.

Although experimental investigations have established many of the chemical species and environmental conditions associated with PEMFC degradation, a holistic mechanical and atomistic understanding remains challenging. Modelling degradation within a PEMFC requires tools that span multiple size and time scales. Specifically, chemical/physical degradation arises from both atomistic and holistic processes.[14] The interatomic processes (including bond cleavage and radical reactions) at the ångström and femtosecond level are just as important as the mesoscale morphological changes.

Computational modelling methods provide a critical framework for bridging these scales by enabling controlled, atomistically resolved studies of degradation phenomena that are difficult to isolate experimentally. By explicitly resolving molecular interactions, reaction pathways, and interfacial structure under well-defined conditions, computational approaches complement experimental observations and offer mechanistic insights into how degradation propagates into structural weakening and eventual delamination.

## 1.2. Density Functional Theory (DFT) and Molecular Dynamics (MD)

Density Functional Theory (DFT) and ab initio molecular dynamics (AIMD) enable highly accurate atomistic simulations of electrochemical processes occurring at the smallest relevant length scales in PEMFCs (see **Table 1**). In particular, DFT provides direct access to electronic structure, reaction energetics, and charge transfer processes. As such, DFT is an indispensable tool for studying surface reactions on platinum catalysts and degradation pathways within polymeric/oligomeric materials. Within the context of PEMFC, DFT has been extensively applied to investigate radical formation, platinum oxidation/dissolution (see **Figure 2A**), bond breaking pathways in Nafion fragments, and adsorption phenomena at catalyst interfaces.[15]

There are several layers to DFT calculations, depending on the clarity with which the electron cloud is represented. There are five layers of accuracy that are often described within the scope of the

"DFT ladder." General gradient approximations (GGA) are the most common when studying the solid-liquid interface. The GGA approach (methods such as PBE [16]) improves upon its predecessor by accounting for the local electron density and the gradient at which it appears. Therefore, it is more accurate for polar bonding structures and molecular geometries. In addition, these functional areas are also paired with a van-Der Waals correctional functional (such as Grimme D3 [17] to account for the long-range or non-covalent dispersion interactions. More advanced versions have also seen their place in studying the PEMFC degradation mechanisms. Meta-GGA methods (like SCAN [18]) include the orbital kinetic energy density in a GGA functional. Therefore, they are better at handling non-covalent interactions and intermediate-range correlations. Despite the more accurate representation of reality, DFT still requires more resources than an MD simulation.

**Table 1**. Comparison of different modelling and simulation techniques used for PEMFC studies, along with introduction of the corresponding experimental methods.[15] AFM = Atomic force microscopy, AST = Accelerated stress test, EIS = Electrochemical impedance spectroscopy, SANS = Small angle neutron scattering, SAXS = Small angle x-ray scattering. SEM = Scanning electron microscopy, TEM = Transmission electron microscopy, USANS = Ultra small angle neutron scattering, USAXS = Ultra small angle x-ray scattering, WAXS = Wide angle x-ray scattering, XRD = x-ray diffraction.

| Method | Scale | Principle | Scope | Pros | Experimental Equivalents |
|---|---|---|---|---|---|
| Ab-initio MD | 1 Å to 100 nm, fs to 10 ps | Quantum mechanics | Activation energy, potential dependent properties, adsorption geometry, dissolution energy barrier | Quantum mechanic energy accuracy, no force field needed | XRD, NMR, TEM |
| DFT | 1 Å to 100 nm, fs to 10 ps | Quantum mechanics | Interactions at Pt/C catalysts and electrocatalyst activities | First principles accuracy, wide applicability | WAXS, SAXS, USAXS, SANS, USANS, TEM |
| Classical MD | 1 Å to 100 nm, fs to 10 ns | Newton's law of motion and force fields | Ion transport analyses, water diffusion, membrane evolution, and mechanical deformations | Wide range of length and time scale, flexible boundary condition | AFM, SEM |
| Kinetic Monte Carlo | 1 Å to 1 μm, ns to ms | Statistical sampling | Membrane dissolution, degradation mechanisms, and deposition processes | Real time dynamics, flexibility in state transitions definition | Electron microscopy, FTIR |
| Coarse Grained MD | 1 Å to 1 μm, ms to s | Newton's law of motion, Langevin equation, | Analysis of solvent self-assembly behavior and phase separation in Nafion | Extended mesoscale phenomena, lower computational costs, simple | |
| Continuum Models | >100 μm, Ms to ms | Navier stocks transport equations | Impact of compression on PEMFC's performance, flow channel configurations, water and heat transport studies, modeling of stacks, cooling | Well validated methods, advanced physical modeling, high order schemes | EIS, thermal conductivity measurement techniques, current measurements, AST, I-V curves, IR thermography |

Molecular dynamics (MD) simulations provide a powerful framework for investigating atomic and molecular motion over time, enabling direct access to structural, transport, and mechanical phenomena that are inaccessible to purely electronic structure methods. Conventional MD simulations employing non-reactive force fields are widely used to study polymer morphology, water channel formation, ion diffusion, and the mechanical response of materials under stress or hydration cycling.[15]

Reactive MD approaches, most notably ReaxFF,[19] extend these capabilities by allowing bond breaking and formation, enabling explicit simulation of radical attack, side-chain scission, and oxidative degradation processes. Compared to DFT, MD methods are computationally more affordable and can therefore access larger interfacial regions and longer timescales, making them particularly valuable for studying PEMFC relevant interfaces.

Several force fields have been employed in PEMFC related MD studies, including COMPASS,[20] Universal,[21] and DREIDING.[22] Among these, COMPASS is derived from a combination of ab initio calculations and experimental data, although it is known to overestimate thermal conductivity in some polymer systems.[15] Despite this, it has been acknowledged to outperform DREIDING in computational accuracy at understanding the metal membrane interface.[15]

Combining MD with DFT simulations provides a complementary, multiscale description of PEMFC degradation phenomena. Combining Monte Carlo and MD methods leverages complementary strengths, allowing simulations to explore vast system configurations of different particle behaviors, leading to a comprehensive understanding of a system. The image charge method to quantumly describe an adsorbate and treat the surface as a classical system is a well known example.[23] However, care must be taken to understand the strengths and weaknesses of MD and DFT approaches. Specifically, most force fields struggle to predict the weaker and short-range interactions present in adsorption (see **Figure 2B-F**). Due to the complexity of degradation processes, relatively few computational studies address them directly, and often these works involve a multistage approach of varying modelling techniques.

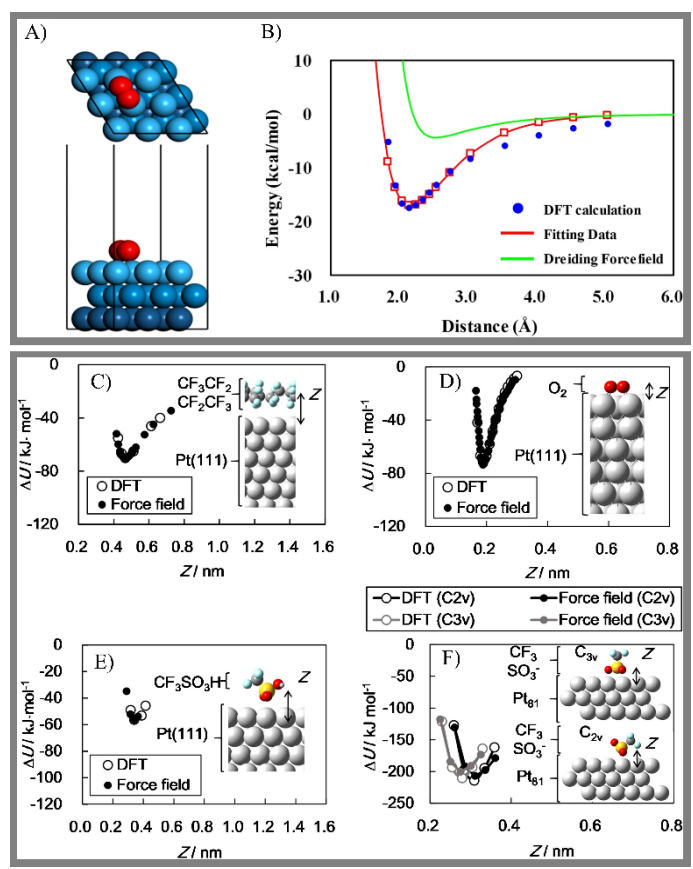

**Figure 2: A)** A simple DFT optimized structure of an O$_2$ molecule adsorbing onto the Pt(111) structure. **B)** The change in potential energy by varying the O$_2$ – Pt distance as determined by DFT and DREIDING force field. There is a large discrepancy between DFT and the force field at the critical bonding distance. **C-F)** The change in potential energy by varying the adsorbate – Pt distance as determined by DFT and COMPASS force field. COMPASS performs significantly better at predicting weaker interactions in a variety of situations. Reused with permission from.[24,25]

## 2. Chemical Degradation Mechanisms

### 2.1. Challenges in Modelling Electrochemical Potentials with DFT

Potential cycling is detrimental to the catalyst layer, leading to gradual oxidation and dissolution of the Pt atoms into the electrolyte. However, DFT previously lacked the necessary tools to manipulate the applied potential simulation. Electrochemical processes at electrified interfaces play a pivotal role in PEMFCs. Yet they are challenging to replicate.[26] The interface is non-crystalline and complex, which proves challenging for periodic simulated systems. The computational resources are significant as an electrified interface involves many ions and solvent molecules across varying timescales.[27] The electrostatics and polarization are critically challenging as they are resource intensive and troublesome to reference the calculated values in a meaningful way due to poor reference data (see **Figure 3**).[28] The interface is difficult to isolate experimentally as it is relatively small compared to the bulk solution/metal. The dynamic nature of the interfacial structure adds a greater challenge as surface adsorption, fluctuating charge and the surrounding solvent all impact the properties of the surface.[29,30] Only recently have computational tools been developed to cost effectively describe the electrified interface. Both constant-charge and constant-potential simulation approaches are viable depending on the system under study.[26] The computational hydrogen electrode is also valuable for verifying computational findings and extrapolating them to real world findings.[31] Despite this, more work is needed in this field as few publications focus on this problem.

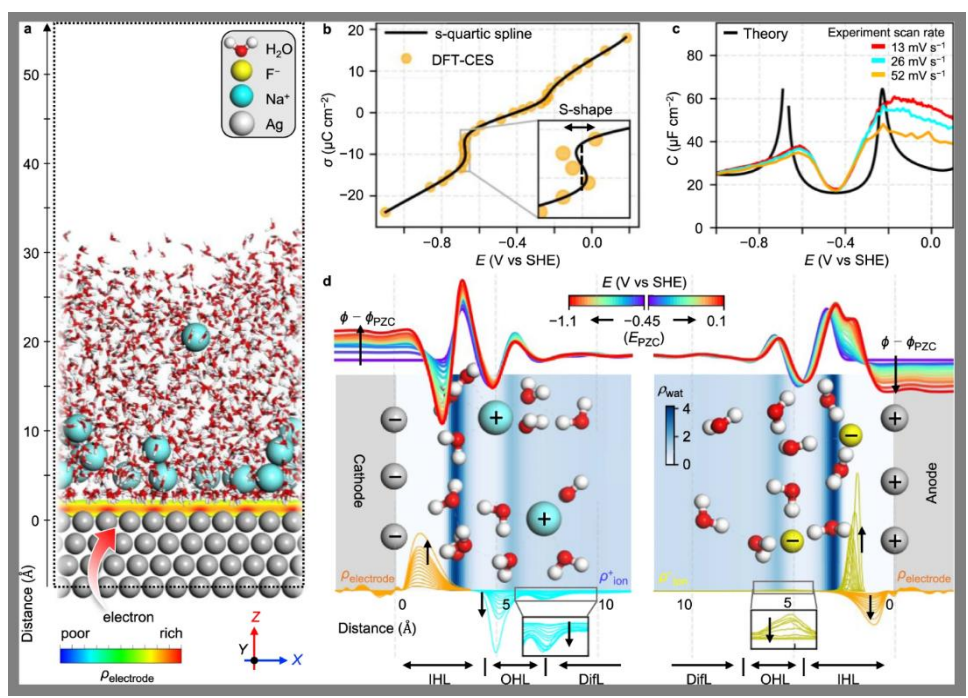

**Figure 3**: **A)** A snapshot of the simulation system consisting of an Ag(111) electrode-electrolyte-vacuum interface by.[27] The excess charge density of the metal electrode, $\rho_{electrode}$, is shown as a color map that is screened by either excess $Na^+$ or $F^-$ ions in the electrolyte. **B)** Surface charge density, $\sigma$, calculated as a function of the electrode potential, $E$ (vs standard hydrogen electrode (SHE)). A dashed vertical line is constructed along the coexisting line of two different charge states. The 'S' shaped curve is challenging to replicate in computational simulations, as it is due to the constant fluctuation of charge close to the potential of zero charge. **C)** Comparison between the camel shaped curve of the differential capacitance, $C$, versus $E$ with that of the experiments on Ag(111) electrode in a 3 mM NaF electrolyte. Another incredibly challenging figure to replicate in simulation is often significantly underestimated in experiments. The challenge originates from several origins. The dynamic nature of the interface makes it challenging to fully account for the water structures.[32] It is also important to note that static or cold models typically yield peak features that are sharper and more intense than those observed in experimental measurements. As a result, capacitance measurements play a critical role in providing reference data for comparison with theoretical models. **D)** Representative structures of hydrated ions near the electrodes. Local density profiles of water, $\rho_{wat}$, are shown as color maps in the background (unit: g cm$^{-3}$) that define the location of the inner Helmholtz layer, outer Helmholtz layer, and Diffusion layer. $\rho_{electrode}$, and the ion charge density profiles, $\rho_{ion}$, are also shown in the below row. Local electrostatic potential profiles, $\phi$, are shown in the upper row, as a function of $E$. The black arrows indicate an increase in negative or positive charging from the PZC.

A common strategy for simulating an applied potential is to either adjust the number of electrons in the system (constant-charge) or vary the Fermi level (constant-potential).[26] In practice, the constant-potential approach more accurately reflects experimental conditions and demonstrates improved scalability, as it better represents the relevant thermodynamic ensembles and reduces finite-size effects.[18] On the other hand, constant-charge calculations exhibit greater resilience against discrepancies, such as deviations in interfacial capacitance and absolute potential alignment.[26] Therefore, the choice between these approaches should be determined by the surface characteristics and charge state rather than by the modelled charge–potential relationship. Consequently, both methods are widely employed in the literature.

## 2.2. Overview of Chemical Degradation

The chemical degradation of membranes is well documented and represents one of the most widely discussed failure mechanisms in PEMFCs.[33–35] The Nafion membrane and ionomers in the catalyst layer are particularly susceptible to chemically driven degradation, especially through radical attack.

The decomposition of hydrogen peroxide ($H_2O_2$) on Pt or through reaction with carbon-centered radicals is highly exothermic, generating reactive hydroxyl (•OH) and hydrogen (•H) radicals that serve as primary initiators of membrane degradation. These radicals preferentially target vulnerable sites in Nafion, such as the tertiary carbon linking the side chain to the main chain, or manufacturing defects like carboxylic acid groups, which drastically lower the energy barrier for ·OH initiated "unzipping" of the polymer backbone (see **Equation 1a**). Consequently, these species are generally present in low concentrations, even under favorable operating conditions. Their formation is most frequently associated with start-up and shutdown cycles or periods of fuel starvation [10]. Under open-circuit voltage conditions, the simultaneous presence of oxygen and hydrogen during these events, particularly in low humidity environments, can lead to the formation of H₂O₂ (see **Equation 1b**).[36,37]

$$H_2O_2 + e^- \rightarrow OH^- + OH \cdot \ |\ H_2O_2 \rightarrow OOH^- + H^+\ |\ OOH^- \rightarrow HOO \cdot + e^-$$ **Equation 1a** [38]

$$O_2(aq) \rightarrow O_2(ads) \xrightarrow{2H^+} H_2O_2(ads) \rightarrow H_2O_2(aq)$$ **Equation 1b**

DFT has been especially valuable for elucidating degradation mechanisms in Nafion, where experimental characterization techniques primarily infer reaction pathways from observed degradation byproducts rather than directly resolving transient intermediates. While these approaches have yielded a deep understanding of radical induced degradation, it remains widely accepted that although Nafion defect sites are the primary targets for radical attack, degradation can still proceed in chemically pristine Nafion molecules.[34] The C–S bond within the sulfonic acid functional group is particularly susceptible to cleavage due to its relatively low reaction barrier. This degradation pathway disrupts proton transport and can ultimately lead to mechanical failures such as delamination. Understanding these degradation mechanisms is critical, as the breakdown of either Pt or Nafion initiates a feedback cycle that accelerates degradation at the interface. The extremely short lifetimes of reactive radicals and intermediates make these reactions challenging to capture experimentally, rendering first-principles simulations the most effective tool for mechanistic insight.

## 2.3. Radical-Induced Degradation Pathways

Free radicals attack the polymer backbone and side-chain structure, resulting in the loss of SO₃⁻ groups and a decrease in the tensile properties, water absorption, and proton conductivity of the catalyst-coated membranes.[33,39] However, it is understood that the free radicals do not limit the gas permeability of oxygen and hydrogen.[40] The ionomer embattlement will reduce the elasticity of the membrane until it breaks away under sufficient mechanical stress. Oxidation stress can also occur on the catalyst layer (CL) due to radicals (see **Equation 2a**).[38,41] The radicals attack the CL, causing corrosion of the carbon carriers and thinning the structure (see **Equation 2b**).[42]

$$OH \cdot (aq) + Pt \rightarrow Pt - OH \cdot + OH \cdot (aq) \rightarrow H_2O(l) + PtO_2$$ **Equation 2a**

$$OH \cdot (aq) + C(graphite) \rightarrow CO(aq) + H \cdot (aq)$$ **Equation 2b**

The radicals involved occur in several forms, each possessing distinct properties that enable them to attack different components of the PEMFC.

## 2.3.1. Hydrogen Radical (H•) and Hydroxide (OH•) Mechanisms

Initially, Bajaj, A. et al. (2020) investigated the degradation pathways of both pristine and defect containing Nafion molecules.[43] A key finding was the identification of the hydrogen radical (H•) as a dominant agent in initiating degradation of pristine Nafion chains (see **Figure 4**). Hydrogen radicals are abundant during fuel cell operation but have previously been proposed primarily as secondary degradation agents, with limited in situ experimental validation.[44]

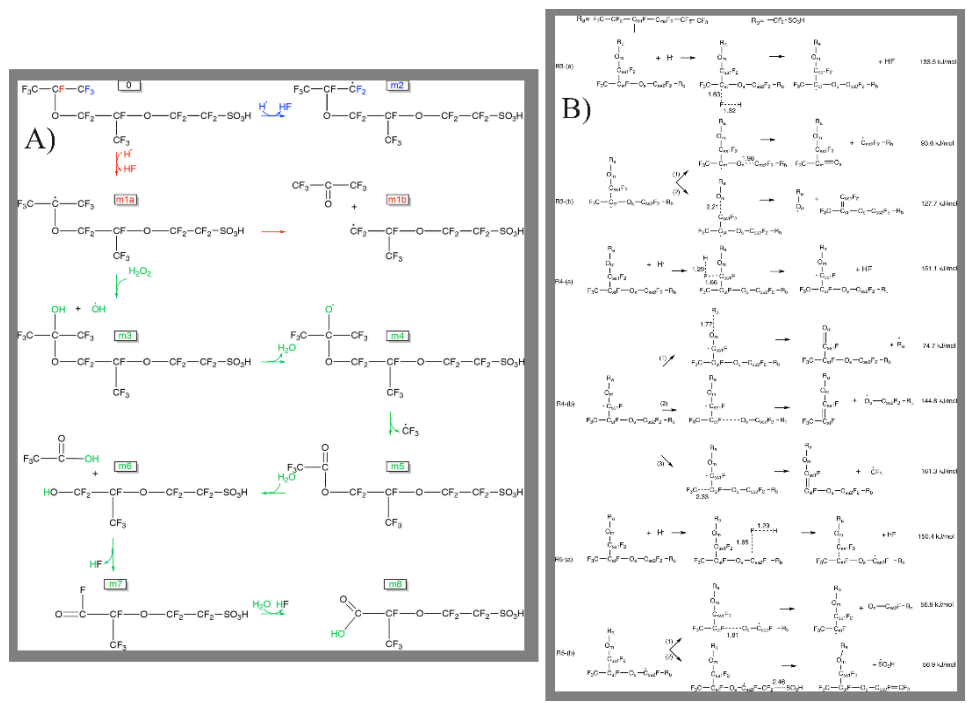

**Figure 4: A**) H radical induced attack mechanism on the backbone of the Nafion molecule, reused with permission from.[43] Blue arrows and intermediate labels denote attack initiating at the terminal carbon site, and red arrows and intermediate labels correspond to attack initiating at the tertiary carbon site, followed by degradation. The green arrows correspond to intermediate and reaction steps that occur when $m_{1a}$ reacts with $H_2O_2$. The green $H_2O_2$ mechanisms are highly favorable and are energetically downhill. **B**) The proposed degradation mechanisms for side chain scission, selected atomic distances in transition states (in Å), and reaction barrier heights for H• radical attacks on $C_{mt}$ and $C_{ms}$. Reused with permission from.[44]

Both studies demonstrated that H• radicals induce degradation more efficiently than hydroxyl (OH•) radicals. While OH• preferentially cleaves the weak C–S bond in the sulfonic acid group, H• readily removes fluorine atoms from tertiary or primary carbons on the polymer backbone, forming HF (see **Figure 5**). Importantly, degradation pathways initiated by H• exhibit lower activation barriers than OH• initiated reactions across both main-chain and side-chain carbon sites. Furthermore, H• induced reactions generate activated sites that facilitate subsequent OH• attack, indicating that realistic degradation modelling must explicitly account for the synergistic effects of both radical species.

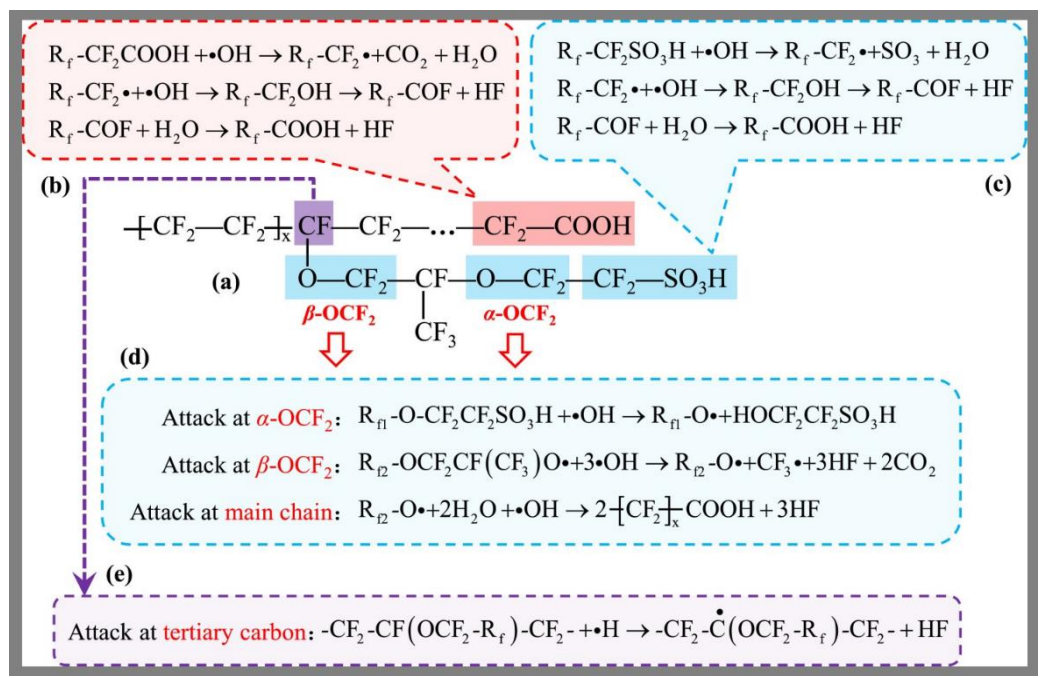

**Figure 5**: Summary of chemical degradation of Nafion. **A)** Molecule structure of a Nafion membrane with long side chains. Chemical degradation mechanisms via free radical attack by **B)** the carboxyl group at the terminal of the main chain, **C)** the C-S bond at the side chain, **D)** α-$OCF_2$ and β-$OCF_2$ within the side chain, and **E)** the tertiary carbon branch point between the main chain and the side chain. Reused with permission from.[39]

### 2.3.2. The Hydronium Ion ($H_3O^+$) and Radical ($H_3O$•)

The role of the hydronium ion remains controversial, with ongoing debate regarding its relative dominance at solid–liquid and liquid–air interfaces. Macroscopic experiments suggest that the air–water interface is basic based on the detection of negative charges at the interface that indicates the enrichment of hydroxides ($OH^-$), whereas microscopic studies mostly support the acidic air–water interface with the observation of hydronium ($H_3O^+$) accumulation in the top layer of the interface.[45,46] Recent studies have revealed that both $OH^-$ and $H_3O^+$ preferentially accumulate near interfacial regions, although their distributions occur at different interfacial depths.[45,47] $OH^-$ would occupy closer to the metal surface, and $H_3O^+$ would sit in the bulk solution due to the charge of the Pt under applied potentials.

More recently, attention has shifted toward the hydronium radical ($H_3O\bullet$), another reactive species presumed to exist under PEMFC operating conditions. The hydronium radical is exceptionally difficult to isolate experimentally due to its extreme reactivity and rapid decomposition into water and hydrogen atoms. Hydronium itself is increasingly mobile, able to share the H• to neighboring $OH^-$ and $H_2O$ species through the bulk solution (see **Figure 6A-F**). Its formation was only recently discussed in the context of the fundamental reaction $H_3O^+ + OH^- \rightleftharpoons 2H_2O$.[48] Owing to its unstable nature, $H_3O\bullet$ has often been neglected in Nafion degradation studies and was previously invoked only as a speculative explanation for the limited sulfur containing degradation products observed experimentally. Its presence would be impossible to detect in experiments, as it leads to the same adduct product as H•.

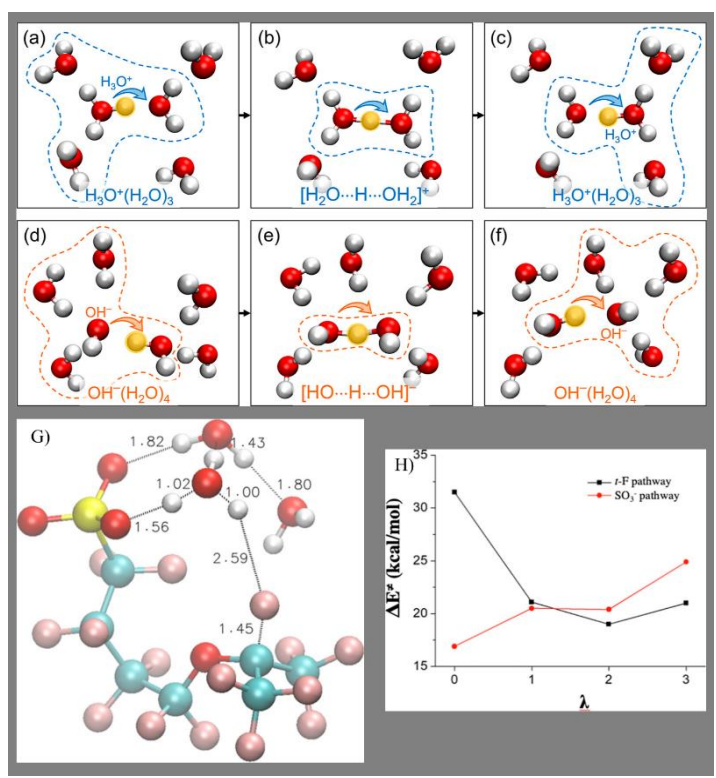

**Figure 6**: **A-F**) Schematic diagrams of hydronium and hydroxide diffusion using DPMD simulations of slab systems. **A-C**) Hydronium and **D-F**) Hydroxide proton migration processes. Protons participating in the transfer process are highlighted in yellow, with arrows indicating the direction of self-ion diffusion. Retrieved with permission from.[48] **G**) The Transition state structure of a simplified Nafion molecule supported by hydronium and water cluster. **H**) Change in potential energy depending on the relative humidity (max 3 water molecules) of the Nafion hydronium complex. t-F refers to the tertiary Fluorine degradation mechanism forming HF with the transition

state shown in **Figure 6G**. The $H_3O•$ and waters bend the shape of the Nafion molecule until a t-F is removed. Reused with permission from.[49]

Long, H. et al. (2022) provided the first detailed DFT-based investigation of hydronium radical induced Nafion degradation.[49] Their work showed that although isolated $H_3O•$ formation is energetically unfavorable, the radical can exist as a stabilized cluster with surrounding water molecules, typically $H_3O•(H_2O)_3$. Moreover, $H_3O•$ was forming a stable intermediate through interaction with the sulfonate ($SO_3H$) group of Nafion. The proposed degradation mechanism involves a conformational bending of the Nafion molecule that brings the hydrophobic backbone into close proximity with the sulfonic acid side chain, thereby enabling the formation of HF or $HSO_3^-$ (see **Figure 6G-H**). Once formed, $H_3O•$ initiated degradation exhibits lower activation barriers than H• induced pathways. Although still in its early stages, this work provides critical insight into the complexity of radical driven Nafion degradation under realistic hydrated conditions. As evident, more work should be focused on the intermediate steps and expand the range of studied radicals.

## 3. Overview of Catalyst Layer Degradation Mechanisms

Degradation of the catalyst layer (CL) follows several pathways: catalyst ageing, particle movement, catalyst overflow, electrolyte dissolution, and carbon coarsening. These effects all result in diminished electrochemically active surface area and catalytic efficiency. Pt nanoparticles undergo dissolution under high potential and redeposition under low potential, a process exacerbated during start–stop cycles and load transients (see **Figure 7A-D**).[8] The degradation of the CL is inherently complex, and isolating the individual mechanisms responsible is difficult because key operating parameters, including temperature, current density, liquid water content, and relative humidity, vary spatially within the FC.[50]

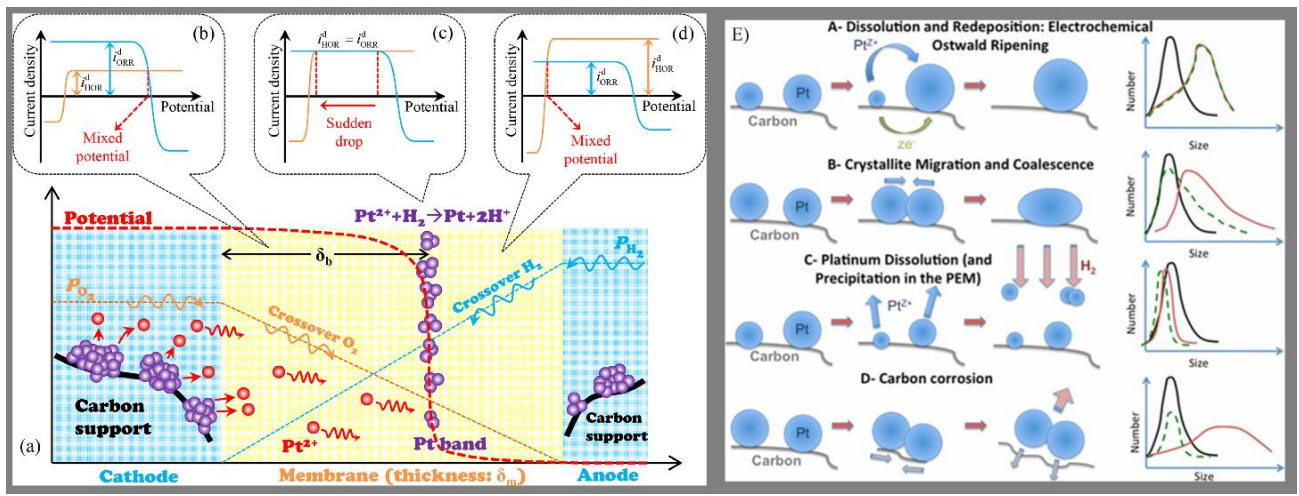

**Figure 7**: **A**) Schematic diagram of the formation of dissolved Pt particles in the membrane and bulk solution impacted by varying potential changes. The effect of mixed potentials during HOR and ORR on the cathodic side **B**), for the Pt dissolved in bulk solution **C**), and on the anode side **D**). $i_{HOR}^d$ is the diffusion limited current density for HOR. $i_{OOR}^d$ is the diffusion limited current density for ORR. Reused with permission from.[39] **E**) Schematic of the typical degradation mechanisms proposed for PEMFC CL degradation. The black histogram represents the initial material, the green histograms represent spherical particle sizes, and the red histograms represent agglomerates. Retrieved with permission from.[51]

The sintering and enlargement of Pt particles over the course of operation are attributed to several mechanisms. One such mechanism is the dissolution of Pt particles with smaller dimensions and their redeposition onto particles with larger dimensions, a process known as Ostwald ripening that results in the growth of the particles (see **Figure 7E**). Ostwald ripening is the precipitation of Pt in the ionomer membrane caused by the dissolution of Pt particles in the ionomeric phase due to the reduction of the Pt ion in combination with the transition of hydrogen from the anode side. The resulting new formation leads to a reduction in membrane longevity and a decrease in ionic conductivity. Agglomeration of Pt can occur at the nano and atomic scale. The formation of larger agglomerated Pt particles occurs through random collisions and differences in surface energies. However, it is not clearly understood which mechanism causes the atomistic agglomeration of atomic Pt. A recent MD study focusing on Al atoms shared that the degree of oxidation of the host Pt nanoparticle can change the behavior of the small conjoining particle.[52] Fundamentally, the large and small particles are competing for the same limited oxidizing resources when they approach. This results in a slowing down of the oxidation surrounding the impact and reduces the final oxygen consumption and lower valence states of Al. This is only the start and the behavior of more hydrophobic metals is still missing.

## 3.1. Potential-Dependent Oxide Formation and Dissolution

Under realistic operating conditions involving high potentials, the formation of stable $PtO_2$ phases further suppresses electrochemical activity. At potentials higher than ~1.1 V vs RHE, a two-dimensional oxide layer composed of $PtO_4$ forms on top of the Pt surface.[53] The $PtO_4$ structure behaved similarly to $Pt_3O_4$, which was reported in other experimental works. Upon cathodic polarization, $Pt_3O_4$ units were reduced to soluble $[PtOH(H_2O)_3]^+$ complexes, stripping Pt atoms from the surface as $Pt^{2+}$ species (see **Figure 8A-C**). Determining the oxidation state of the metal during oxidation is often crucial for understanding the underlying reaction mechanisms and the corresponding energy barriers.[53] DFT has elucidated this atomistic dissolution mechanism, for example through the reaction $Pt_3O_4 + 8H^+ + 6e^- \rightarrow [Pt(H_2O)_4]^{2+} + 2Pt$, which describes oxide decomposition, Pt release, and partial redeposition.

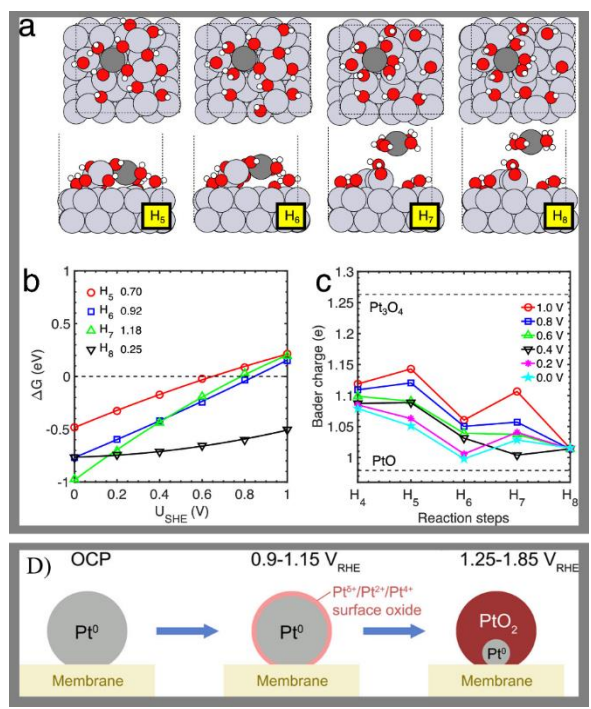

**Figure 8**: Reduction of $Pt(OH)_4$ unit in Pt surface oxide. **A)** Top and side view of the atomic structure of progressive dissolution of Pt. **B)** Potential-dependent reaction free energies of the $Pt(OH)_4$ reduction; the values indicate the number of electrons transferred in the reaction. **C)** Bader charges of the Pt atom in $Pt(OH)_4$ along the reaction coordinate of the $Pt(OH)_4$ reduction at various applied potentials. Retrieved with permission from [53]. **D)** Schematic representation of the electrochemical oxidation of platinum at varying potentials. OCP stands for open circuit voltage. Reused with permission from.[54]

The surface oxide phase forms as part of a wider progression of oxidation states that develop with increasing applied potential. Chemisorbed oxygen species and thin surface oxides can emerge at potentials as low as ~0.85 V vs RHE, typically arising from adsorbed O or OH species prior to the development of stable oxide layers.[55] Experimental in situ XPS and x-ray absorption studies also indicate that the initial electrochemical oxidation yields mixed oxidation states ($Pt^0$, $Pt^{2+}$, $Pt^{4+}$) in the surface layer, with oxide formation onset shifting depending on particle environment and substrate interactions (see **Figure 8D**).[54]

As the potential increases further (> ~1.3–1.4 V vs RHE), bulk like $PtO_2$ (often hydrous $PtO_2 \cdot nH_2O$ under electrochemical conditions) becomes more stable and grows thicker, eventually consuming much of the metallic Pt beneath and forming a passivating layer. The relevant potentials are also dependent on nanoparticle composition, as oxide formation on clean and pristine particles generally requires higher applied potentials.[56] Furthermore, dissolution and redeposition processes have been suggested to occur at potentials exceeding ~0.85 V.[56]

### 3.1.1. Self-Limiting Nature of Pt Dissolution and Recovery

Importantly, platinum oxidation and dissolution are self-limiting processes. Dissolution proceeds through the formation of $Pt(OH)_4$ complexes that require a corner sharing oxide framework.[53] Once the ligands are removed, the neighboring Pt atoms become reduced, thereby suppressing localized feedback dissolution. Priyadarsini, A. et al. (2025) further expanded this understanding by modelling Pt segregation in a solvated environment across a wide potential range.[57] They identified the equilibrium potential between aqueous and solid Pt at 1.37 V and demonstrated that bulk-to-surface Pt segregation beyond this potential is kinetically hindered. Consequently, oxidative degradation is limited during steady state operation but is reactivated during start and stop cycling, which replenishes surface vacancies and restarts dissolution. These vacancies can function as highly active sites that facilitate $H_2O_2$ formation and the subsequent generation of radicals.[58] However, the extent of platinum oxidation is strongly dependent on crystal facet and surface morphology. Different corner sharing geometries can alter ligand coordination and oxidation energetics.[59] Under oxidative stress, corner rich nanoparticles undergo rapid warping, driven by the high surface energy and reduced coordination of corner atoms. Spheres or cuboctahedra are the most resilient to this and serve as an extra barrier for continued Pt mass reduction.

Variations in voltage uniformity are strongly linked to fuel cell degradation. Precise control of the temperature and humidity can limit the effect of voltage variation, but it is impractical for real applications. Although, it is noted that in fuel cells, voltage recovery occurs after short rest periods.

The atomistic mechanisms underlying this phenomenon remain largely unclear. It is generally proposed that excess water accumulation leads to blockage of the porous medium, an effect that may be alleviated during resting periods.[8] Additionally, the recovery observed during degradation may result from favorable changes in the CL microstructure, potentially relating it to surface deformation induced degradation processes.[60]

### 3.1.2. MD Studies of Pt Dissolution

The catalyst layer was investigated using MD simulations, focusing on the dissolution and agglomeration behavior of Pt nanoparticles. In acidic PEMFC environments, acidic species can indirectly drive these processes through the generation of reactive oxidative species.[61] The sintering behavior of supported nanoparticles is closely related to the particle size, temperature, metal-support interaction, and other factors (see **Figure 9A**). Within a fuel cell, the maximum temperature is insufficient for excessive migration.[62] Surface oxidation and size differences in neighboring Pt clusters were often the first step to dissolution for a Pt nanoparticle.[62]

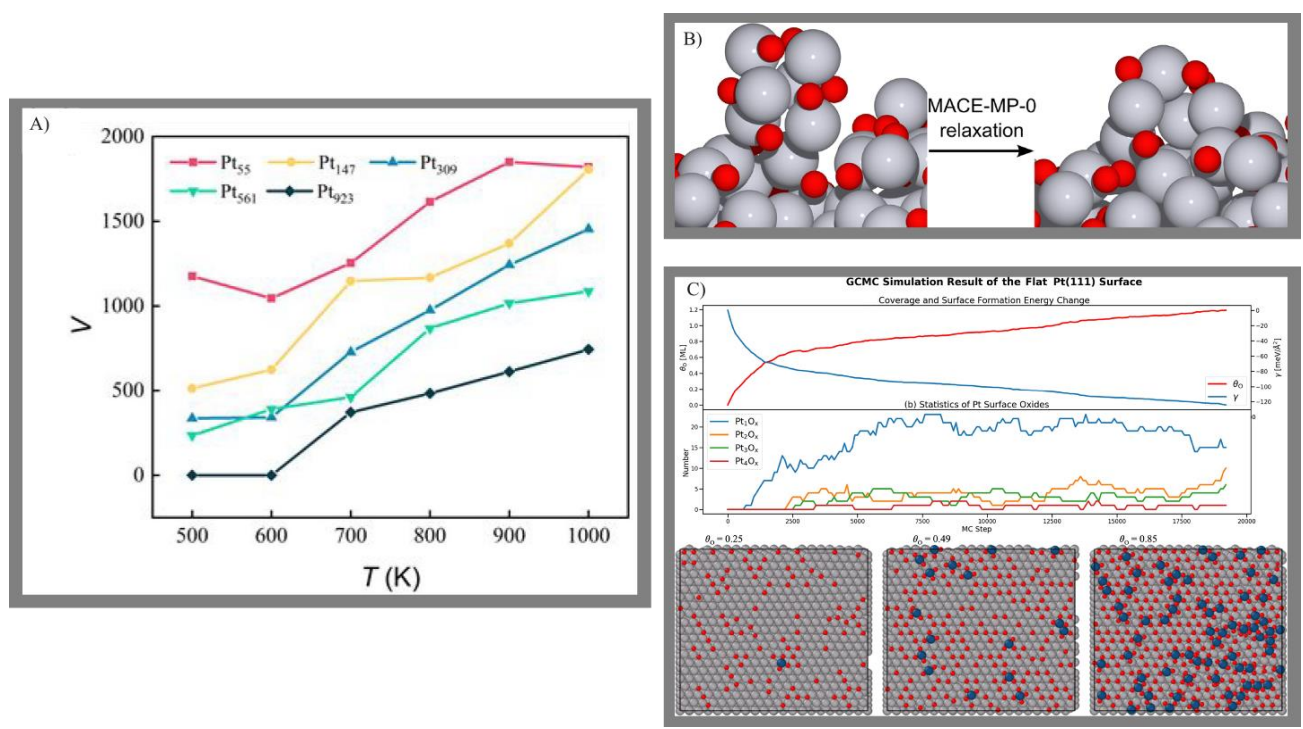

**Figure 9**: **A**) Sintering rate of the same particle on a graphene substrate with variations in temperature. Reused with permission from.[62] **B**) Illustration of the commonly observed phenomena during a MACE-MP-0 geometry optimization, the $Pt_6O_8$ species is packed back into the nanoparticle. Reused with permission from.[63] **C**) GCMC simulation results of the flat p(20 × 20) Pt(111) surface are illustrated. Panel (**A**) shows the changes in surface formation energies and O coverages during oxidation. Panel (**B**) shows the change in the number of local surface oxide units,

such as the square planar $PtO_4$ unit and minimal $PtO_2$ stripe, which are two surface oxides responsible for the $PtO_2$ stripe formation. Panels (**C**) are representative structures at selected coverages. Raised Pt atoms are shown as blue spheres. The first raised Pt atom appears at 0.25 ML coverage, while subsequent Pt atoms occupy FCC adsorption sites. At 0.49 ML coverage, numerous square planar $PtO_4$ oxide units are observed. At 0.85 ML coverage, the surface becomes disordered, and extended $PtO_2$ stripes containing three or more Pt atoms form. Reused with permission from.[64,65]

Oxidation and dissolution of Pt atoms occur rapidly at monolayer (ML) coverages below 0.5, while dissolution beyond 1 ML is reported to be nearly absent.[65] A reduced resistance to dissolution has also been observed with increasing particle size. For larger particles, the formation of a PtO film becomes more probable, as the greater number of supporting Pt atoms can stabilize O substitution. Additionally, smaller nanoparticles would contain high surface energy Miller surfaces (such as 110), which are more reactive and likely to dissociate, revealing the stabler 111 or 100 surface.

Slapikas, R. E. et al. (2023) used classical MD simulations to identify relationships between degradation rate and size/morphology of the nanoparticle.[64] By studying Pt nanoparticles with sizes ranging from 1.35 to 2.92 nm under different hydration levels and temperatures, degradation was observed only at 450 K. This process was attributed to the surface formation of $Pt_6O_8$, which is lifted into the bulk solution owing to its hydrophilicity [66] (See **Figure 9B-C**). After the formation of $Pt_6O_8$, the dissolution mechanism is irreversible, and the Pt is destined to migrate to the bulk solution. However, the existence of the $Pt_6O_8$ structure is debated.[63] There is also an argument that the dissolution mechanism is caused by water itself under cathodic potentials.[64] More recently, Demeyere, T. et al. (2025) reported that, although their results showed excellent agreement with experimental measurements, conflicting binding energies were still observed for the $Pt_6O_8$ species.[63] In fact, comparison with DFT calculations supported MACE-MP-0 reliability, and, along with the lack of experimental evidence for such species, suggested that the $Pt_6O_8$ clusters were likely artifacts from the ReaxFF forcefield.[63] Here, we find that original methodologies are insufficient or are inaccurate at modeling the dissolution process at a large scale. These large scales are imperative as they allow for the exploration of unique yet ever present CL morphologies within the PEMFC. Understanding this oxidation property is dependent on surface morphology, and designing better structures can enable longer lasting CL.[59]

## 3.2. Machine Learning Applications for Catalyst Degradation

Machine learning (ML) potentials have emerged as a transformative methodology for bridging the gap between classical force fields and ab initio simulations, allowing systems to be studied at time scales and accuracy that are beyond the reach of either approach alone.[67] Despite this potential, applications specifically targeting Pt degradation in PEMFCs remain limited in the literature, representing a significant gap given that ML tools are uniquely well suited to replicate the complex multi-body interactions governing catalyst behavior. Recent work has demonstrated that ML potentials can accurately describe the dissolution and synthesis of Pt and Pt-alloy nanoparticles up to 4 nm, capturing the influence of size, shape, and atomic composition on durability.[68] These simulations confirmed that larger particles generally exhibit lower dissolution rates, consistent with well-established experimental trends.[69] Core–shell nanoparticle architectures were further shown to offer enhanced structural stability relative to both ordered and disordered configurations.[68] However, the enhanced stability of core-shell structures comes at a cost to catalytic activity, and whether this trade-off is acceptable remains an open question, particularly given that particle morphology may exert a greater influence on overall performance than core composition.[59]

Understanding the Pourbaix diagrams of realistic nanoparticle systems is also critically important, yet practically intractable for conventional DFT due to the system sizes involved. ML algorithms offer a route to expedite this process, enabling the simultaneous treatment of thousands of atoms. The BE-CGCNN algorithm, for example, enables prediction of bond energies and adsorbate coverages across nanoparticle surfaces.[70] The approach remains in an early stage of development and is currently limited to O and OH surface interactions due to dataset constraints. Yet, extending the method to more complex adsorbate environments would make it an indispensable tool for studying surface oxidation dynamics under realistic PEMFC operating conditions.

## 3.3. Carbon Support Degradation

The structure of the carbon material is deeply intertwined with the performance of the catalyst.[71] The carbon support also protects the Pt nanoparticle from detaching and migrating to other Pt clusters. As carbon is removed (via reacting and turning into $CO_2$ or other products), the Pt becomes exposed and gains high surface energy. To stabilize, it redistributes to another well-embedded Pt cluster.[72] Diffusion barriers for Ostwald ripening are lowered by high electric potentials, elevated temperatures, and humidity within the catalyst layer.[73] These factors facilitate the dissolution of Pt and the formation of a mobile species, and when recombining, increase the average particle size. Consequently, fuel cell efficiency is directly diminished due to the loss of active catalytic surface area. The mechanistic picture derived from DFT indicates that $PtO_2$ migrates either along the support structure or through the gas phase before being reabsorbed by larger particles.[74,75] However, structural anomalies or corrosion of the support structure can induce these defects faster.

### 3.3.1. Carbon Corrosion Mechanisms

Carbon support corrosion leads to particle detachment and loss of connectivity. Carbon corrosion is a term to specifically describe the loss of the carbon support on platinum particles.[8] Carbon corrosion is primarily driven by two mechanisms: transient conditions during startup and shutdown cycles, and localized fuel starvation at the anode under operating conditions.[76,77] The first mechanism arises from a heterogeneous distribution of hydrogen and oxygen gases at the interface, remaining after cell startup or shutdown. The second is unclear. Although fuel shortages can originate from several factors, one of the simplest mechanisms is the localized blockage of portions of the active surface caused by ice formation.

Fuel starvation leads to local hydrogen depletion, resulting in a negative anode potential that promotes the oxidation of water and carbon. Almost all FC components can be physically damaged by repeated freezing and thawing.[78,79] Damage rates depend on local physical conditions and the water distribution in the FC system. These can be more challenging to model as the states of water (frozen or liquid) have no viable method to accurately predict when one starts and another begins in the presence of a metal/polymer. Additionally, the water distribution within the fuel cell can be expensive and often outside the scope of traditional simulations. As of writing, no model has explored the effect of a local dry point phenomena close to the polymer or catalyst layer that would be present due to a fuel shortage.

DFT has been instrumental in understanding Pt–carbon support interactions. Although individual Pt–C interactions are weak and dominated by van der Waals forces, the cumulative interfacial area results in substantial binding energies.[80] However, as the dispersion interaction is non-directional, the nanoparticles likely retain a degree of mobility on the surface (see **Figure 10A**). Nanoparticles with structural defects or high surface energy, particularly those exposing high-index Miller surfaces, may become partially embedded in the carbon support.[81] Therefore, carbon and metal degradation can occur through multiple mechanisms, depending on particle size, dissolution behavior, and detachment processes.[82]

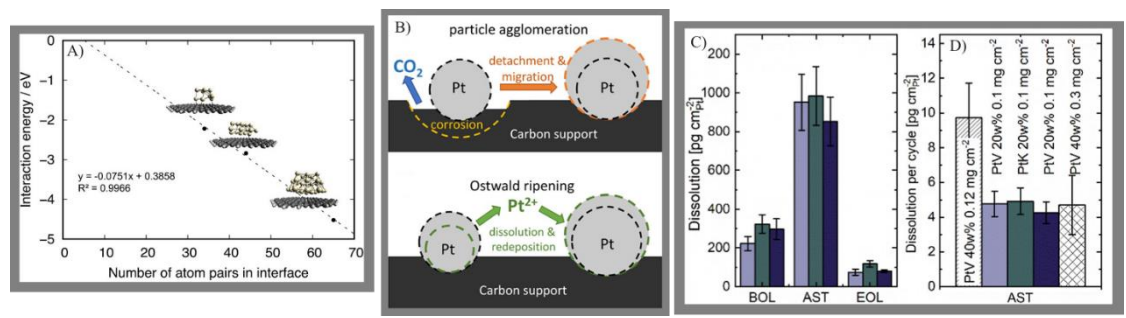

**Figure 10**: **A**) Correlation between Pt-C interaction energy and number of atom pairs in the interface. Reused with permission from.[80] **B**) Schematic representation of the modeled particle growth mechanisms on carbon support. The corrosion of the surrounding particle will eventually lead to detachment from the carbon support and consequential migration. Otherwise, the Pt particles dissolution occurs more slowly via the migration of smaller $Pt^{+2}$ clusters. The difference in surface area between varying sized nanoparticles is the driving mechanism. Reused with permission from.[72] **C-D**) The Pt dissolution results for Pt Ketjenblack (Pt/KB) and Pt Vulcan (Pt/C) at performance optimized ratios of carbon to Pt. **C**) Summary of dissolution results for beginning of life (BOL), accelerated stress test (AST), and end of life (EOL), normalized by specific area of the Pt. **D**) Comparison of the obtained result with dissolution results of different CL thicknesses. Retrieved with permission from.[83]

The early stages of degradation are governed by particle size growth through Ostwald ripening; however, degradation arising from dissolution and detachment progresses linearly with time [82] (see **Figure 10B**). Presumably, the surface energy of the Pt particle will only allow it to grow so large before it breaks off into the solution. The choice of carbon support is also critical, as traditional nonporous supports such as Vulcan exhibit lower tolerance to losses in electrochemically active surface area and shorter operational lifetimes than highly porous supports such as Ketjenblack [83] (see **Figure 10C-D**). The porous architecture of these materials suppresses nanoparticle coalescence by restricting particle migration within the carbon pore network. As a result, the confinement effect increases interparticle spacing and stabilizes the catalyst nanoparticles against agglomeration.[84,85]

### 3.3.2. Defects and Structural Anomalies in Carbon Supports

The quality of the carbon support is crucial for preventing CL degradation. On defective graphite supports with large vacancies, there are stronger Pt-C binding strengths [86] (see **Figure 11A**). These sites would likely form during a radical attack within the fuel cell.[87] Studies have also indicated that $SO_4$ groups formed under harsh operating conditions may play an active role in puncturing carbon supports.[88] These holes have strong binding energies that attract neighboring Pt or PtO particles.[87]

As the size of the hole increases, the detachment energy barrier for Pt NP decreases. For large Pt nanoparticles, detachment energies decrease as the particle splits and migrates toward energetically favorable defect sites, leaving residual clusters behind. Solvation of the Pt species and graphite defects further lowers the detachment barrier, with defect-anchored water molecules assisting the gradual detachment process.

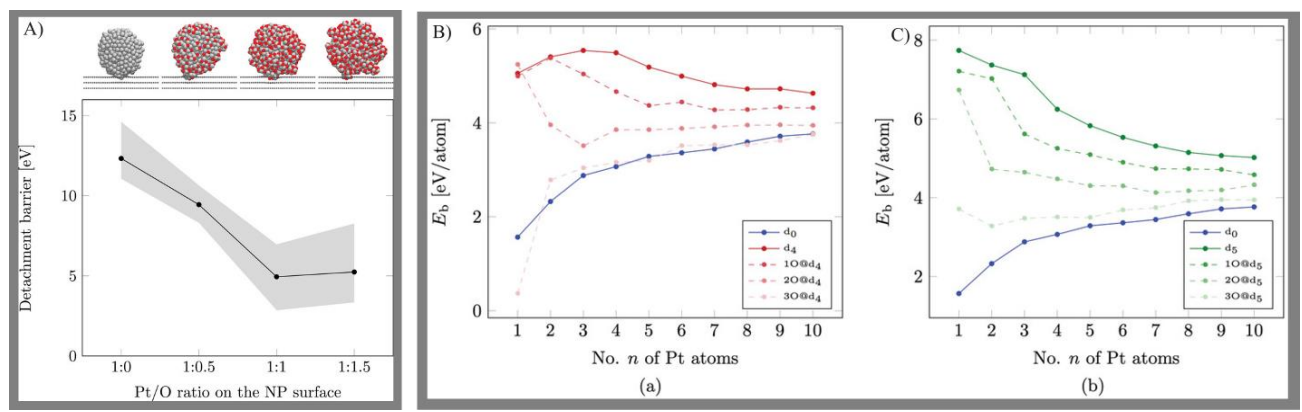

**Figure 11**: **A**) Detachment barriers of 3 nm Pt nanoparticles as a function of Pt/O surface ratio. The gray area indicates the min/max range of the energy barriers. Reused with permission from.[86] **B-C**) Comparison of the binding energy of Ptn clusters on gradually oxidized carbon supports **B**) $d_4$ graphene supports (red dashed lines) and **C**) $d_5$ graphene supports (green dashed lines) with the growth of $Pt_n$ clusters on $d_0$ and $d_4$ or $d_5$ graphene sheets (solid lines). Reused with permission from.[89]

Structural defects and anomalies in fuel cell components fundamentally dictate the stability and degradation pathways of catalysts in PEMFCs. Vacancy defects in graphitic carbon supports act as critical anchoring points that significantly increase the binding strength of platinum nanoparticles compared to pristine, defect free surfaces.[89] (see **Figure 11B-C**). These defects exert a templating effect, which limits the size of platinum clusters and helps stabilize small nanoparticles against migration and coalescence.[89] However, these same defects can become liabilities under harsh operating conditions; for instance, the expansion of vacancy defects during high-potential cycling (1.0–1.6 V) promotes the kinetics of the carbon oxidation reaction (COR).[90] Continued progression of this process destabilizes the support structure, culminating in its collapse and the shedding or swarm like aggregation of the catalyst.[90]

Chemical anomalies, such as oxygen containing functional groups and ionomer degradation products, further accelerate performance loss. Oxygen containing defects on graphene supports weaken the platinum–carbon interaction and generate an oxygen rich environment that promotes platinum oxidation and stabilizes higher oxidation states.[89] This environment intensifies Ostwald ripening, particularly during normal load cycling. Conversely, vacancy defects may provide a protective electronic effect by lowering the platinum d-band center, which reduces its susceptibility to oxidation and consequently suppresses dissolution during potential cycling.[90]

### 3.3.3. Degradation Under Different Operating Conditions

PEMFCs operating under idle conditions and load cycling experience distinct physical, chemical, and electrochemical degradation mechanisms that limit durability.[91] In particular, open-circuit voltage conditions promote platinum dissolution and migration, while also accelerating chemically driven membrane degradation. High potentials initiate the oxidation and subsequent dissolution of Pt particles into the ionomer phase. These dissolved Pt ions migrate through the membrane and are reduced by crossover hydrogen from the anode, forming a "Pt band" within the membrane that can compromise its stability.[91]

Meanwhile, during dynamic cycles, changes in current density cause rapid fluctuations in heat and water production. These fluctuations induce hygrothermal cycling, causing the membrane to repeatedly swell and shrink, which leads to mechanical fatigue, micro-cracks, and delamination between the catalyst layer and the membrane.[8] During voltage transients or startup-shutdown events within cycles, the cathode can experience potential overshoots as high as 1.5 V. This causes the carbon support to oxidize into $CO_2$, leading to particle detachment, structural collapse of the catalyst layer, and increased mass transport resistance. Frequent potential cycling inhibits the formation of a stable, protective oxide layer on Pt particles, leading to dissolution rates that are substantially higher than those under steady-state conditions. Fundamentally, dynamic cycling creates favorable conditions for multiple degradation pathways by inducing heterogeneous distributions of temperature, humidity, and fuel, while also promoting radical formation.[92–94]

### 3.4. Machine Learning Applications for Carbon Support Modeling

The application of machine learning to carbon support degradation represents one of the most computationally demanding frontiers in PEMFC modelling, as it requires capturing reactive bond breaking events across large systems and long timescales. Active machine learning models have begun to address this challenge for carbon nanostructure growth and degradation on metallic surfaces. Zhang, D. et al. (2024) developed an active ML model combining Gaussian Approximation Potentials with molecular dynamics and time stamped force biased Monte Carlo methods to simulate the full dynamic growth of graphene on Cu(111).[95] Their approach accurately reproduced key subprocesses including carbon monomer diffusion, ring and chain formation, and edge-passivated growth as well as bond breaking events induced by ion impact. Critically, by extending the simulations to other metal surfaces including Cr(110) and Ti(001), the framework demonstrated transferability across substrate types, a prerequisite for modelling Pt-decorated carbon supports where the metal-carbon interface is chemically heterogeneous. This work establishes a methodological template for studying how carbon nanostructures nucleate, grow, and degrade under the influence of metallic substrates relevant to PEMFC catalyst layers.

The interfacial chemistry between oxidized carbon supports and water is equally important, as oxidation of the carbon surface governs both Pt anchoring and local proton transport. Azom, G. et al. (2025) employed deep potential molecular dynamics (DPMD) simulations with deep neural network force fields to investigate proton transfer at the graphene oxide-water interface.[96] By comparing graphene oxide sheets at two different oxidation levels, they demonstrated that oxidized graphene oxide promotes interfacial proton release, whereas reduced graphene oxide instead adsorbs protons. This oxidation dependent switching of interfacial proton behavior has direct implications for PEMFC carbon support degradation: as the carbon support oxidizes progressively during potential cycling, the local proton environment at the Pt-carbon-ionomer triple-phase boundary shifts in a manner that could alter both catalytic kinetics and ionomer adhesion. The use of DPMD here is particularly significant, as it captures Grotthuss-type proton transfer with near-DFT accuracy at system sizes and timescales inaccessible to conventional ab initio methods.

More broadly, ML driven electrocatalyst design has been comprehensively reviewed by Ding, R. et al. (2024), who surveyed ML applications across hydrogen evolution, oxygen evolution, hydrogen oxidation, and oxygen reduction reactions.[67] Within this framework, graph neural network approaches have proven especially powerful for identifying complex correlations between carbon support structure, including defect density, graphitization degree, and heteroatom doping, and electrocatalytic performance descriptors. These methods enable high throughput screening of carbon support configurations that would be prohibitive to evaluate individually with DFT, providing a route toward rational design of supports with optimized Pt anchoring strength and corrosion resistance.

Despite these advances, several gaps remain. Current ML potentials for carbon systems are predominantly trained on pristine or lightly oxidized graphene, and their accuracy under the high-potential, acidic, and hydrated conditions characteristic of PEMFC operation has not been systematically validated. The coupled degradation of the carbon support and the Pt nanoparticle, where carbon oxidation lowers the Pt detachment barrier and increases Ostwald ripening rates has not yet been modelled within a single ML framework. Generating the diverse, high quality DFT reference datasets required to train transferable potentials for these multicomponent, reactive interfaces remain the critical near-term challenge.

## 4. Impurities and Contaminants

Contaminants are among the most critical concerns for fuel cell technologies (see **Figure 12**). Either introduced during the construction procedure or through fuel intake, they accelerate degradation through poisoning and side reactions. Therefore, it is a constant effort to ascertain the consequences of contamination and how to recognize their presence. Generally, impurities can be classed into two categories: gas diffusion layer contaminants and catalyst layer breakers. Density Functional Theory (DFT) studies have elucidated that while Pt(111) is intrinsically the most stable crystal plane, its surface integrity is significantly compromised under operating conditions by the specific adsorption of oxygen reduction reaction (ORR) intermediates (∗O, ∗OH, ∗OOH) and electrolyte ions. Atomic oxygen adsorption weakens the binding energy of surface Pt atoms more than hydroxyl or hydroperoxyl species, but electrolyte anions such as $PO_4^{3-}$ and $SO_4^{2-}$ exert a far greater destabilizing effect.[97] Cations have also been demonstrated to impact the degradation and performance of PEMFCs.

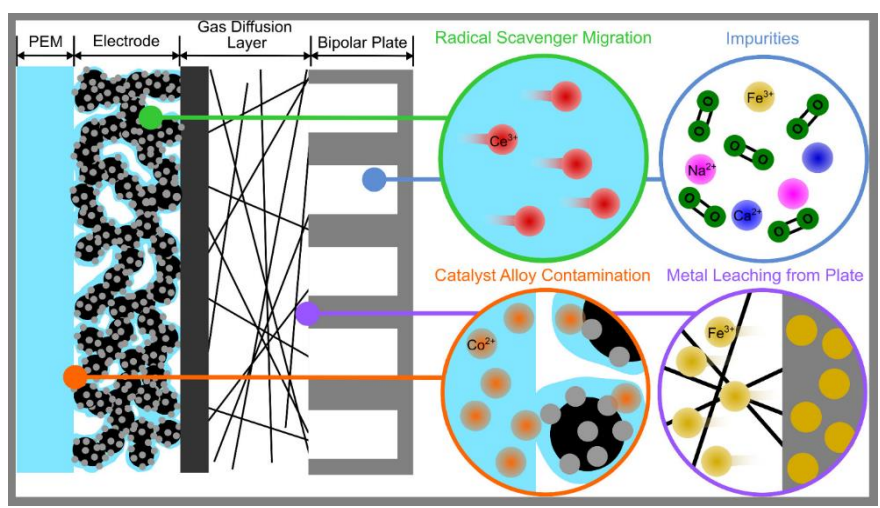

**Figure 12**: Schematic summarizing different sources of cation contamination for PEMFCs. Cation contamination can originate from multiple sources and locations within the PEMFC, including radical scavenging cerium from the membrane, impurities introduced from environment, the alloying element of the catalyst leaching, and corrosion of metallic components. Reused with permission from.[98]

## 4.1. Mechanisms of Cation Damage

Although cation contamination in PEMFCs is often discussed in terms of individual metals, the underlying degradation pathways can be more rigorously classified into four mechanistic regimes: (i) electrostatic crosslinking dominated, (ii) hydration dominated, (iii) redox driven chemical degradation, and (iv) interfacial electrochemical modulation. This framework enables a clearer comparison of $Ca^{2+}$, $Al^{3+}$, $Cu^{2+}$, $Mg^{2+}$, and $Fe^{2+}/Fe^{3+}$ and reveals important gaps in current simulation studies (see **Table 2**).

**Table 2**: Summary of cation contaminates.

| Cation | Molecular Level Impact | Typical Source of Contaminant | Ref |
|---|---|---|---|
| Na | Competes with $H^+$ for sulfonate sites; lowers proton concentration; perturbs water channel connectivity; modifies EDL contributing to Pt instability. | Water/air contamination; leaching from gaskets, seals, metallic parts. | 99–101 |
| K | Competes with $H^+$ for sulfonate sites; weaker hydration than $Na^+$; modifies cathode interfacial structure and ORR kinetics. | Water/air contamination; introduced via handling or manufacturing processes. | 99 |
| Ca | Bridges sulfonate groups; increases ionic crosslink density; narrows hydrophilic channels; stiffens membrane; disrupts H-bond network; impairs HOR and ORR. | Water/air contamination; sealants; electrolyte impurities; system materials. | 99,101,102 |
| Mg | Strong hydration shell; limited crosslinking vs $Ca^{2+}$; sequesters water; reduces H-bond rearrangement; suppresses proton mobility; possible salt precipitation. | Water/air contamination; magnesium containing materials (gaskets, seals). | 99,101,102 |
| Fe | Catalyzes •OH formation via valence cycling; induces membrane chemical scission; strong sulfonate coordination; accelerates Pt dissolution and voltage loss. | Leaching from metal alloys; corrosion products; airborne contamination. | 103,104 |
| Cu | Valence cycling enables ROS formation; anisotropic coordination distorts ionomer nanostructure; modifies catalyst-layer EDL; alters ORR kinetics and durability. | Leaks from copper components (wiring, system parts); copper dust contamination. | 105 |
| Al | Extremely strong hydration; slow ligand exchange; strong sulfonate binding; sequesters water; reduces proton mobility; limited chemical degradation. | Aluminum containing components (seals, metallic parts). | 104,106 |

$Ca^{2+}$ represents the prototypical electrostatic contaminant. Its divalent charge enables strong bridging between sulfonate groups, effectively increasing ionic crosslink density within Nafion.[107] This restricts polymer segmental motion, narrows hydrophilic channel widths, and reduces proton conductivity. Simultaneously, $Ca^{2+}$ forms complexes with $H_3O^+$ and $H_2O$,[101] disrupting hydrogen-bond connectivity and increasing ohmic resistance. The primary degradation mechanism here is therefore structural: conductivity loss emerges from constrained morphology rather than radically induced chain scission. Mechanical stiffening effects, reflected in modest increases in elastic modulus,[108] further support a crosslinking dominated interpretation. Importantly, $Ca^{2+}$ is environmentally prevalent, explaining why $Ca^{2+} \gg K^+ > Na^+$ in practical severity.[109]

While $Mg^{2+}$ is often grouped with $Ca^{2+}$ as a divalent alkaline earth contaminant, its physicochemical behavior within hydrated Nafion domains is fundamentally distinct. In practical systems, $Mg^{2+}$ is environmentally prevalent in feed water and humidification systems, often coexisting with $Ca^{2+}$ [110]. The combined presence of both ions may lead to cooperative or competitive binding effects within ionic clusters. Yet most computational investigations consider them independently, overlooking potential multi-ion coupling phenomena. Owing to its smaller ionic radius and higher charge density, $Mg^{2+}$ exhibits substantially stronger hydration than $Ca^{2+}$, forming tightly bound first hydration shells with slow water exchange kinetics.[111] This difference has important mechanistic consequences.

Like $Ca^{2+}$, $Mg^{2+}$ can bind to sulfonate groups and compete with protons for ionic sites, contributing to reduced proton concentration and conductivity.[108] However, its stronger hydration shell limits direct inner-sphere coordination to multiple sulfonate groups simultaneously. In contrast to $Ca^{2+}$, which readily forms electrostatic bridges and increases crosslink density,[107] $Mg^{2+}$ is more likely to remain partially solvated when interacting with ionic domains. Consequently, its structural impact may manifest less through permanent crosslinking and more through local water immobilization and reduced hydrogen bond network flexibility.

From a transport perspective, $Mg^{2+}$ introduces a distinct "hydration rigidity" regime.[111] It's strong binding to water reduces the population of dynamically exchanging molecules that sustain Grotthuss proton hopping.[112,113] Rather than collapsing water channels via crosslinking, $Mg^{2+}$ may decrease proton mobility by reducing water reorientation rates and hydrogen-bond rearrangement dynamics. This distinction is subtle but important: $Ca^{2+}$ primarily restricts polymer motion and narrows channels, whereas $Mg^{2+}$ may preserve channel geometry but dampen internal water dynamics.

Unlike $Fe^{2+}/Fe^{3+}$ and $Cu^+/Cu^{2+}$,[114] $Mg^{2+}$ is redox inactive under PEMFC operating conditions and therefore does not directly catalyze hydroxyl radical formation. Its degradation mechanism is therefore purely structural and transport mediated. This raises an unresolved question: does $Mg^{2+}$ reduce conductivity more through suppressed water mobility than through polymer crosslinking? Existing simulation studies have not systematically compared residence times, coordination numbers, or water diffusion coefficients for $Mg^{2+}$ versus $Ca^{2+}$ within equivalent membrane environments.

$Al^{3+}$ introduces a distinct regime governed by extreme charge density and strong hydration. The $[Al(H_2O)_6]^{3+}$ complex exhibits slow ligand exchange kinetics and strong water binding, limiting both mobility and redox participation.[99] Unlike Fe or Cu, $Al^{3+}$ cannot undergo valence cycling and therefore does not directly catalyze Fenton type radical formation [115]. Its impact is instead mediated through water sequestration and strong electrostatic interaction with sulfonate groups.

From a molecular transport perspective, $Al^{3+}$ likely reduces the fraction of dynamically exchanging water molecules available to sustain Grotthuss proton hopping. However, its strong hydration shell may also spatially confine its perturbation to localized domains. Thus, $Al^{3+}$ presents a mechanistic paradox: high electrostatic binding but low chemical reactivity. Whether its net impact is transport limiting or morphologically reorganizing remains insufficiently quantified. Current MD studies have not systematically compared binding free energies or hydration free energies of $Al^{3+}$ versus $Ca^{2+}$ within realistic membrane environments a significant gap given their distinct hydration thermodynamics.

In contrast to $Ca^{2+}$ and $Al^{3+}$, $Fe^{2+}/Fe^{3+}$ and $Cu^+/Cu^{2+}$ introduce a chemically reactive regime. Their ability to cycle between oxidation states enables Fenton type hydroxyl radical production,[115] directly accelerating membrane chemical degradation. Fe, in particular, strongly interacts with sulfonate groups and has been incorporated into cross-linked systems,[116] while side-chain functionalization can stabilize $Fe^{3+}$ and promote radical formation.[117] This dual structural and chemical role explains the severe voltage losses observed experimentally.[104]

Cu shares similar redox flexibility but introduces additional complexity due to anisotropic coordination geometries and intermediate hydration strength. Cu contamination affects both membrane chemistry and catalyst-layer structure.[105] Cu operates through coupled structural distortion, redox chemistry, and interfacial modification. Even at relatively low concentrations, such redox active species may dominate long-term degradation due to catalytic amplification effects.

Cations also influence catalyst durability through modification of the electric double layer (EDL). Pt dissolution kinetics are strongly governed by local electric fields and cation hydration shells.[118] Smaller cations such as $Li^+$ can enhance Pt dissolution by altering interfacial hydroxide concentration, [119] which neutralizes dissolved $Pt^{2+}$ species and prevents redeposition. Here, degradation is neither purely structural nor radically driven, but electrochemically mediated. The acidity and hydration chemistry of specific cations further modulate this process.

Importantly, these regimes are not mutually exclusive. For example, $Cu^{2+}$ combines redox activity with structural coordination, while $Ca^{2+}$ influences both crosslink density and water network connectivity.[101] The dominant degradation pathway likely depends on concentration, hydration level, and local electrochemical environment. Yet most computational studies isolate single mechanisms or single contaminants, limiting predictive capability.

Despite extensive experimental investigation into inorganic contaminants,[108] simulation based studies remain sparse and mechanistically fragmented. Few studies quantify comparative binding free energies, diffusion coefficients, hydrogen-bond network metrics, or local electric field perturbations across different cation classes within the same membrane model. Without such systematic comparisons, it remains unclear whether degradation severity is primarily governed by charge density, hydration free energy, redox potential, or interfacial field effects.

A unified computational framework that evaluates electrostatic binding strength, hydration thermodynamics, redox accessibility, and interfacial field coupling within a consistent molecular model would enable predictive ranking of contaminant severity. Such an approach would move beyond descriptive categorization toward mechanistically resolved, quantitatively grounded durability modelling of PEMFC systems.

## 4.2. Anion and Gaseous Contamination

Anionic contaminants ($Cl^-$, $SO_4^{2-}$, $F^-$, $SO_3^-$, $CO_3^{2-}$) differ fundamentally from cationic species (see **Table 3**). They do not primarily disrupt bulk proton transport via ion exchange; instead, they exert their influence at electrode electrolyte interfaces. Their impact is therefore more directly coupled to catalytic kinetics, interfacial electronic structure, and local acid base equilibria.[120–122] In contrast to simple site blocking models, these species often induce coupled electrochemical chemical degradation loops that accelerate both catalyst and membrane failure. In addition, PEMFCs are highly susceptible to performance degradation from gaseous contaminants such as $CO$, $CO_2$, $NH_3$, and $H_2S$, primarily through catalyst poisoning and changes in ionomer chemistry.

**Table 3**: Summary of anion contamination.

| Contaminant | Impact on Degradation | Source of Contaminant | Ref |
|---|---|---|---|
| $Cl^-$ | Competitive adsorption on Pt active sites; suppression of ORR kinetics; ligand-assisted Pt dissolution; displacement of sulfonic acid groups reducing proton conductivity; catalyst thinning and increased charge-transfer resistance | Leaching from gaskets/seals; impurities in water supply; environmental exposure (e.g., saline environments) | 123–125 |
| $SO_4^{2-}$ / $SO_3^-$ | Strong chemisorption on Pt step/kink sites; severe ORR suppression; alteration of Pt electronic structure; largely irreversible catalyst poisoning; acceleration of carbon support corrosion | Airborne sulfur species; fuel impurities; degradation of Nafion ionomer producing $SO_3^-$ fragments | 120–122,126 |
| F | Indicator of membrane chemical degradation | Decomposition of perfluoro sulfonic acid membrane under oxidative stress | 127 |
| $CO_3$ | Formation of insoluble salts in membrane; reduction in proton conductivity; disruption of ion transport channels; increased ohmic resistance | $CO_2$ contamination from ambient air; reaction of $CO_2$ with water forming carbonic acid species | 121,128 |
| CO | Strong adsorption on Pt; inhibition of HOR (anode) and ORR (cathode); voltage losses at ppm levels; increased overpotential leading to Pt dissolution during mitigation | Reformate hydrogen streams; trace impurities in supplied hydrogen fuel | 129–132 |
| NOx | Adsorption on Pt active sites; formation of nitric acid at high potentials; membrane hydrolysis and oxidative degradation; increased radical generation | Air or oxygen feed contamination; industrial emissions; environmental pollution | 133–135 |
| SOx | Sulphur poisoning of Pt; severe suppression of HOR/ORR; largely irreversible deactivation; enhanced carbon corrosion | Air contamination; sulfur-containing fuel impurities | 136 |
| $H_2S$ | Dissociative adsorption forming strongly bound sulfur species; near complete catalyst deactivation; surface reconstruction; carbon support corrosion | Hydrogen fuel impurities (reformate systems); ambient contamination | 132 |
| $NH_3$ | Neutralization of sulfonic acid groups; reduction in ion exchange capacity and proton conductivity; increased ohmic losses | Contamination in fuel or air streams; agricultural or industrial emissions | 92–94 |

Fluoride ions are typically not externally introduced but arise from chemical degradation of the perfluoro sulfonic acid membrane. Radical induced scission of –$CF_2$– backbone structures generates $F^-$ species, making fluoride release a widely used diagnostic metric for membrane chemical degradation.[127]

Although $F^-$ exhibits comparatively weak adsorption on Pt, its significance lies in its correlation with membrane weight loss, thinning, and pinhole formation. Elevated fluoride release rates indicate enhanced oxidative stress and peroxide formation, often preceding catastrophic membrane failure.[127] Thus, fluoride functions primarily as an indicator of degradation severity rather than as a direct catalyst poison.

$Cl^-$ is among the most pervasive anionic contaminants due to its presence in processing water, sealing materials, and environmental exposure. Mechanistically, $Cl^-$ chemisorbs strongly onto Pt surfaces, particularly at low-coordination sites such as edges and corners, which are also the most active sites for the oxygen reduction reaction (ORR).[121,137] Competitive adsorption between chloride and oxygenated intermediates shifts ORR onset potentials negatively, increases charge-transfer resistance, and reduces apparent exchange current densities.

Beyond site blocking, chloride accelerates Pt dissolution under potential cycling conditions by stabilizing soluble Pt–Cl complexes ($[PtCl_4]^{-2}$).[138,139] This promotes nanoparticle thinning and

Ostwald ripening, thereby coupling adsorption poisoning with structural catalyst degradation. Simultaneously, chloride disrupts the ionomer phase by displacing sulfonic acid groups, reducing proton conductivity and altering the local acidic environment.[125] These dual effects of surface poisoning and ionomer destabilization render chloride contamination particularly detrimental in humidified PEMFC operation.

Sulfur containing anions represent a more severe poisoning pathway. Sulfate ($SO_4^{2-}$) and sulfonate ($SO_3^-$) species including fragments derived from Nafion degradation bind strongly to Pt surface.[80] These species preferentially occupy step and kink sites, which are critical for optimal ORR activity. Their adsorption strength exceeds that of oxygen and key ORR intermediates,[80] significantly altering the Pt d-band structure and suppressing catalytic turnover. Sulfur poisoning proceeds more rapidly and is less reversible than carbon based contamination. Removal generally requires high potentials that further accelerate Pt dissolution and carbon support corrosion.[126] Consequently, sulfur contamination does not merely reduce active surface area but fundamentally destabilizes electrode architecture.

Carbonate and bicarbonate species originate from $CO_2$ dissolution in humidified operating environments. $CO_2$ forms carbonic acid, which dissociates into bicarbonate and carbonate anions that can accumulate within the membrane phase. These species form insoluble salts and partially obstruct hydrophilic proton conducting channels, reducing effective proton conductivity and altering water management.[140] In contrast to sulfur poisoning, carbonate effects are often partially reversible if $CO_2$ exposure is eliminated. Their primary impact lies in ionic transport disruption rather than strong surface chemisorption.

Carbon monoxide (CO) is a well-established catalyst poison, particularly at the anode. CO binds strongly to Pt via σ-donation and π-backbonding interactions,[141] forming stable Pt–CO complexes that inhibit oxygen and hydrogen adsorption even at trace concentrations.[142,143] Recovery typically requires high overpotentials or air bleeding strategies, both of which accelerate catalyst degradation. Compared to sulfur species, carbon based contaminants are generally more reversible, although repeated mitigation cycles exacerbate Pt dissolution and carbon corrosion.[129]

Ammonia acts primarily as a proton scavenger. Reaction with sulfonic acid groups forms $NH_4^+$ species, reducing ion exchange capacity and proton conductivity. Prolonged exposure results in persistent membrane conductivity loss and structural degradation.[92–94] Unlike CO or sulfur species, $NH_3$ predominantly disrupts membrane transport chemistry rather than directly poisoning catalytic surfaces.

# 5. Mechanical Degradation and Interfacial Delamination

Mechanical degradation is thoroughly examined and has a key impact on fuel cell lifetime. This mechanism presents a particular challenge for computational modelling because the relevant failure events, span timescales many orders of magnitude beyond what atomistic simulation can directly access. Mechanical stress within the membrane can induce cracks, pinholes, tears, and punctures, ultimately leading to reduced performance. These failures produce microscopic structural changes that promote delamination as well as membrane swelling and shrinkage under operating conditions. Fundamentally, the Nafion membrane swells when hydrated and contracts when dried, producing tensile and compressive stresses. This repetitive stress weakens the membrane. These defects compromise gas separation and accelerate chemical attack by providing diffusion pathways for radicals and reactants. Often these issues are caused by improper assembly of the MEA or defects in the material.[144,145] There is a well-established relationship between relative humidity, mechanical stress, and temperature as key operating factors contributing to membrane fatigue.[146]

Humidity cycling and tensile loading are particularly damaging during PEMFC startup and shutdown cycles. MD and multiscale simulations indicate that cracks readily form in the CL and can propagate into interfacial delamination.[147] Due to the complexity of these processes, relatively few computational studies address them directly. However, very few studies have focused on this property, as it is challenging to model the startup/shutdown cycles. Often, these works involve a multistage approach of varying modelling techniques.

## 5.1. Temperature Effects

The temperature is a key variable towards impacting the output and efficiency of PEMFCs. The membrane and gas layers are sensitive to changes in temperature, and changes beyond the operating range can impact gas diffusivity and membrane.[148] If the temperature becomes too high, the membrane's internal resistance increases as liquid water evaporates, forming water vapor that raises the internal pressure within the membrane structure. Conversely, when the temperature is too low, condensed water accumulates within the cell, leading to flooding that obstructs ion transport pathways. Therefore, effective regulation of both water content and temperature is essential for maintaining fuel cell performance and ensuring long-term device durability (see **Figure 13A-C**). The fuel cell stack is designed to tolerate complex mechanical stresses; however, the membrane resistance increases with applied pressure.[149] Therefore, homogeneous changes within the fuel cell are less impactful. Yet, heterogeneous pressure variations are more common and can be induced externally (pressure differences between the inlet and outlets) or internally (condensation of water due to poor migration).

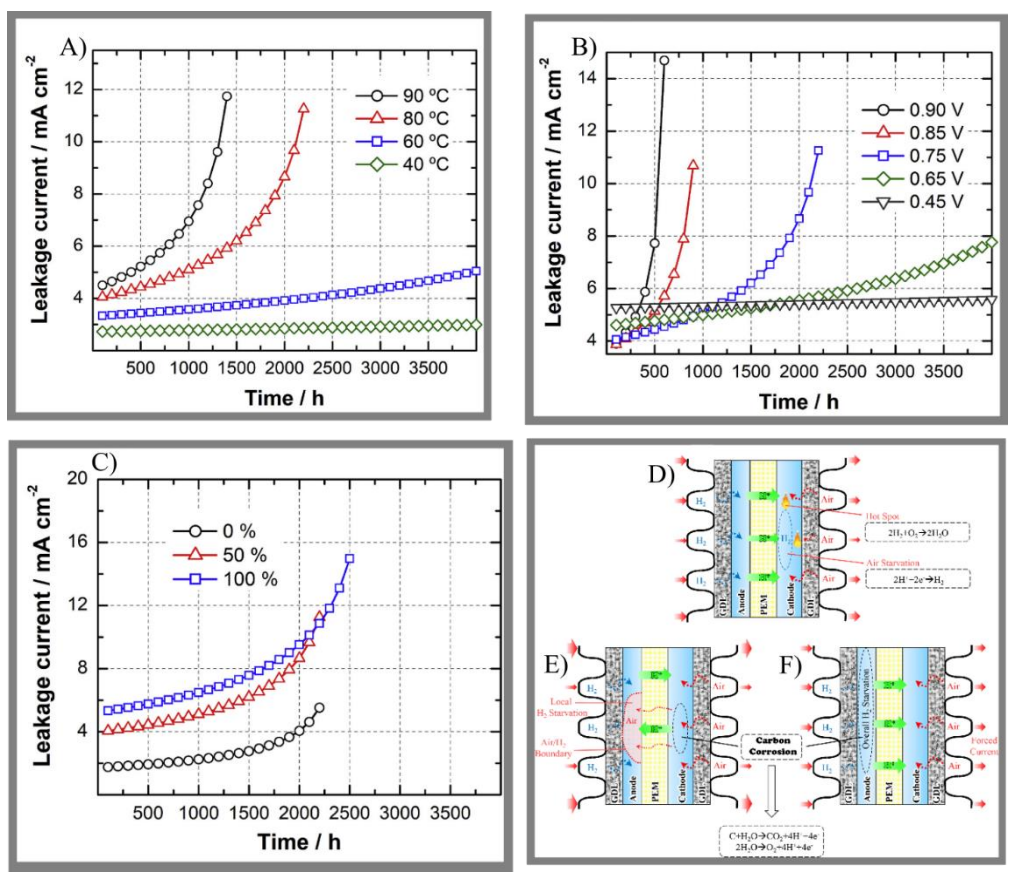

**Figure 13**: **A-C**) Simulation results for leakage current over time as a function of **A**) temperature, **B**) potential and **C**) relative humidity. Leakage current is unwanted current paths within a PEMFC, primarily from hydrogen crossover and internal short circuits. Reused with permission from.[36] **D-F**) Schematic diagrams of **D**) air starvation, (**E**) local fuel starvation, and (**F**) overall fuel starvation in the load changing process. Reused with permission from.[39]

Optimal power output is typically achieved within a relatively narrow temperature window of approximately 30–100 °C, which inherently limits the operational flexibility of the technology. In addition to external thermal conditions, internal temperature increases may also occur due to localized hot spots generated by exothermic reactions or regions of elevated electrical resistance. These localized temperature gradients can significantly increase membrane stress and may ultimately lead to catastrophic failure (see **Figure 13D-F**). Uneven gas distribution and impaired transport properties within the membrane or catalyst layer can further exacerbate hot-spot formation.[39,150] The resulting thermal gradients promote membrane dehydration and expansion, which increases mechanical stress and reduces structural integrity, thereby accelerating membrane degradation.

## 5.2. Humidity and Freeze-Thaw Cycling

A similar effect occurs with humidity. When the membrane is exposed to high humidity during cycles, it induces mechanical stress and disrupts the material's membrane cohesion. Worse effects are experienced under a series of wetting and drying, as the supporting water that sits between the Nafion bundles escapes, leaving the structure to support itself. As a result, membrane degradation becomes more pronounced, with an increased likelihood of mechanical defects such as tears and punctures, making this effect more damaging than temperature fluctuations.[36,151]

Ice formation within PEMFCs during sub-zero operation or freeze–thaw cycling introduces severe mechanical and structural degradation due to volumetric expansion of water upon freezing. The approximately 9% volume expansion generates significant tensile and shear stresses within the hydrated Nafion membrane and catalyst layers, resulting in microcrack formation, disruption of hydrophilic transport channels, and irreversible loss of proton conductivity.[152] Classical MD can make an approximation of this effect but lacks the tools for an accurate representation. Thermal fatigue is accumulative and can act in non-equilibrium ways within a fuel cell (local hot spots).[153,154] Therefore, the relationship between simulated homogeneous temperature changes and experimentally observed localized thermal failure therefore remains poorly established.

Repeated freeze–thaw cycles exacerbate stress accumulation, promoting delamination at membrane–catalyst and catalyst–gas diffusion layer interfaces, which increases interfacial resistance and accelerates performance decay. Additionally, ice formation can trap reactive species and locally concentrate mechanical stress, enhancing stress assisted chemical degradation of the ionomer backbone and side chains. As a result, freeze–thaw damage is a critical durability concern, particularly for automotive PEMFC systems exposed to cold-start conditions. MD simulations have also provided insight into the freezing induced degradation. The crystal surface can impact the nucleation of ice. Surfaces like Pt(111) can provide an ideal framework for ice nucleation to begin.[155] Meanwhile, Pt(211) atomic spacing is too wide and results in weak surface ice structures.[156] Therefore, the ice structure is disordered and limits wettability on the surface. This disorder leads to a stronger delamination force as the ice forms larger, further away from the surface, while forming a small vacuum region between the ice and CL.

## 5.3. Computational Studies of Mechanical Response

Feng, C. et al. investigated the impact of stress concentration regions on the mechanical performance of the CL during tensile loading and humidity cycling.[157] Dehydration induces contraction of the Nafion layer at the Nafion/Pt–C interface, resulting in interfacial stresses that can lead to delamination.[158] Under very humid conditions, the membrane is more malleable and strains under stress. Porosity has a similar effect by reducing membrane density, which increases flexibility but compromises structural integrity. Maximum strain is localized at the Pt particle, meaning that the onset of delamination immediately degrades performance. To mitigate this, an interfacial strength of approximately 12 MPa is required. Another approach is to control the swelling and expansion coefficient of the interface. This value is tied to many factors, including the length of the Nafion and the size and content of Pt in the CL.[159] Failure of either Pt or Nafion reinforces a negative feedback mechanism that hastens interfacial degradation. Limiting Nafion bending and shrinkage is therefore a key strategy for mitigating degradation across wet and dry operating conditions.

### 5.3.1. Hydration and Temperature Analysis

MD simulations offer valuable insight into the mechanical stress responses of polymer membranes. Applying tensile deformation to simulated polymer volumes enables the evaluation of mechanical properties such as elastic modulus and tensile strength, while also revealing how structural perturbations influence these properties. Numerous studies have also explored hydration and temperature dependent effects using MD, frequently demonstrating strong agreement with experimental measurements.[160] Through the use of MD simulations, a concrete model of contraction and expansion of the Nafion membrane was developed and is predictable for a large sample size.

The coordination number (CN) of the first solvated cell around the sulfur group of Nafion was dependent on the temperature and humidity of the cell. The clumping of water molecules limits the mobility of ions and reactants within a fuel cell. However, studies identified that the clumping of water molecules follows a pattern.[160] Water clusters initially formed with a maximum size of ~40 molecules and did not grow beyond this limit across the range of temperatures and humidities studied. These clusters subsequently coalesced into a single megacluster, whose formation was facilitated by humidity, with bulk water molecules acting as bridges that stabilize the Nafion framework. This macrostructure indicates a breakdown of phase separation at high hydration, accompanied by increased intercluster connectivity. Suppressing megacluster formation would help preserve phase separation and sustain Nafion permeability.

With the combined approach of Monte Carlo and MD methods, an understanding of the kinetics and mechanisms of Nafion unzipping was studied.[161] Although slower than main-chain unzipping, side-chain scission is particularly detrimental because it accelerates degradation by functionalizing the

Nafion backbone with carboxyl groups. While fibril filled microvoids can impart strain tolerance in degraded bulk membranes, comparable structural reconstruction is unlikely at the thin interface between the catalyst layer and Nafion.

The catalyst layer is also susceptible to mechanical breakdown. Catalyst migration on the membrane surface further reduces the ductility and mechanical strength. Defect sites and cracks in the membrane collect migrating Pt atoms. These Pt atoms poison and accelerate the decay of the membrane interface. Perforations and pinholes resulting from degradation allow reactants to cross over between electrodes, leading to the mixing of hydrogen and oxygen, unplanned exothermic reactions, and a significant decline in fuel cell performance.[150,162] However, the degradation of the catalyst layer is initiated by locally elevated chemical degradation, leading to potential sites for fracture initiation.[163]

# 6. Interfacial Effects and Transport Phenomena

The interface between the carbon support, Pt nanoparticle, and Nafion ionomer plays a decisive role in PEMFC stability. Degradation often originates at these boundaries, where disparities in mechanical, chemical, and electrostatic properties cause delamination or structural reorganization. The Pt–ionomer interface is particularly critical because it governs both proton transport and catalyst accessibility. Under hydrated conditions, water layers can act as lubricating films, reducing adhesion, while repeated hydration cycles lead to mechanical fatigue.

To date, MD based reviews focusing specifically on PEMFC degradation remain relatively sparse. Most MD studies concentrate on polymer electrolyte membranes typically Nafion or related perfluoro sulfonic acid (PFSA) materials at the molecular scale. Classical and reactive MD simulations bridge atomic and mesoscale descriptions, capturing the structural and dynamic behavior of hydrated membranes and catalyst layers over nanosecond to microsecond timescales.

## 6.1. Oxygen Transport at Nafion-Metal Interfaces

A representative example is the work of Kwon, S. H. et al. (2021).[25] The free energy barriers reported for oxygen diffusion are therefore most reliable under open-circuit conditions and their applicability under cathodic operating potentials is uncertain,[25] which established a detailed molecular framework for understanding hydration and oxygen permeation at Nafion–metal interfaces, building on earlier work.[24] These studies revealed that oxygen molecules preferentially accumulate in interfacial regions between hydrophilic and hydrophobic domains, often near Nafion side chains. At high hydration levels, water molecules form strongly dipolar networks that restrict oxygen mobility, effectively creating a barrier that limits oxygen access to the catalyst surface [24,164] (see **Figure 14**). This behavior is attributed to the organization of the carbon backbone, which defines preferential dissolution pathways for nonpolar species. These simulations often only capture the electrostatically inert Pt surface. The reorganization of the Nafion interface at an applied potential cannot be captured this way. Conventionally, it is extremely challenging, but progress is being made in developing ML interatomic potentials to predict structures.[165] Future work could focus on adding Nafion DFT data to train a new model to understand the Nafion Pt interface at non open circuit potentials.

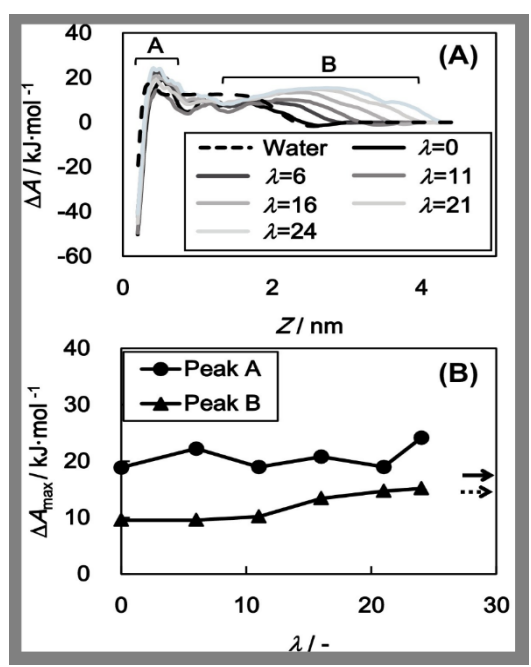

**Figure 14**: **A**) Helmholtz free energy profiles o$A$ of $O_2$ molecules across the Nafion/Pt interface scaled with the free energy at gas phase and (**B**) the heights of peaks A and B shown in **Figure (A)**. Reused with permission from.[24]

At low hydration, the distinction between hydrophilic and hydrophobic domains is less pronounced, resulting in relatively uniform oxygen mobility. In contrast, at high hydration levels, oxygen molecules are displaced from water-rich regions as water–water interactions dominate. The mechanism of oxygen migration is therefore closely tied to Nafion morphology and hydration state. Since Nafion side chains are common targets of radical attack,[166] degradation of these sites are expected to alter oxygen solubility and mobility. However, this relationship has not yet been explicitly investigated using MD simulations and is currently inferred from separate transport and degradation studies.

## 6.2. Water Transport and Proton Dynamics

Water management is a central factor governing PEMFC performance and degradation. Utilizing MD simulations, a better understanding of how water participates and interacts within the interface has been achieved. The diffusion of both $H_2O$ and $H_3O^+$ can be decomposed into three modes: (1) localized binding to $SO_3^-$ groups, (2) confinement within the water channels, and (3) normal diffusion along the tubular pathways. These modes show a hierarchy in the spatial scale and are modulated by hydration. In particular, at low water contents, diffusion is more restricted, and the structural role of hydration is more prominent for efficient $H_3O^+$ transport. The study underscores the critical role of nanometer and nanosecond scale dynamics in governing proton conductivity in PFSA membranes.[167] These findings are also supported by recent work that focused on ionomer content on molecule migration [168] (see **Figure 15**).

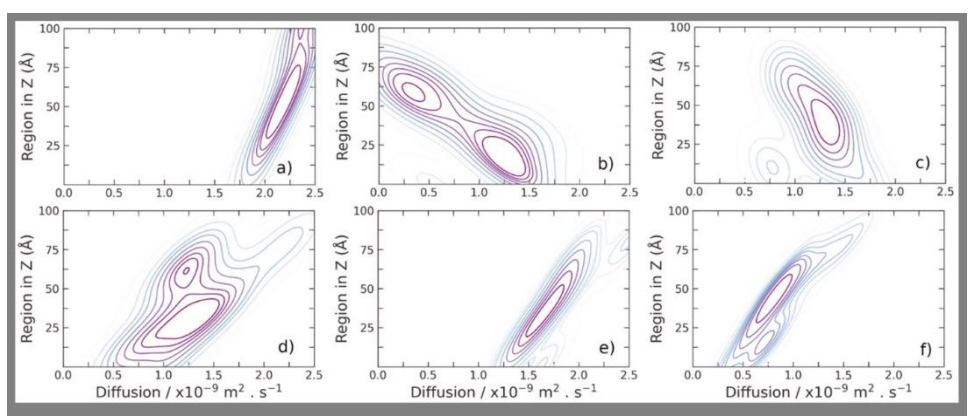

**Figure 15**: Distribution of the average diffusion coefficient of water molecules along the Z-axis in the Water/VC system (**A**), and Pt/VC/Nafion systems with 4, 8, 12, 24, and 36 oligomers (**B, C, D, E, F**). Reused with permission from.[168]

The water channels are a critical function of the Nafion membrane, which allows for the movement of molecules while being safeguarded within the Nafion protective case. These structures form due to the polymer backbone twisting and interconnecting, forming a nanotube-like structure. The hydrophobic backbones are highly organized to prevent contact with water. However, there is an ideal concentration of Nafion oligomers, as too many will restrict water flow due to Nafion's compact nature.

A key mechanism of Nafion degradation was a change in the molecular weight or functionalization. Changing the weight and length of Nafion monomer limits their mobility to form the metal-water-Nafion interface [169,170] (see **Figure 16**). A higher concentration of the sulfonic acid group (removal of the hydrophobic tail) results in tight ionomer packing and limits water mobility. A value of 20-

30% w/w in a Vulcan carbon system exhibits the best performance in terms of water diffusion.[169,170] Therefore, the breakdown of this hydrophobic backbone will result in slower water migration and reduced performance. These models are often simulated with defect-free Nafion with homogenous nature. Meanwhile, real membranes contain side-chain defects produced during synthesis or initial startups.[171,172] While these understanding are valuable, there remains questions on the conditions and structures of late stage fuel cell operation.

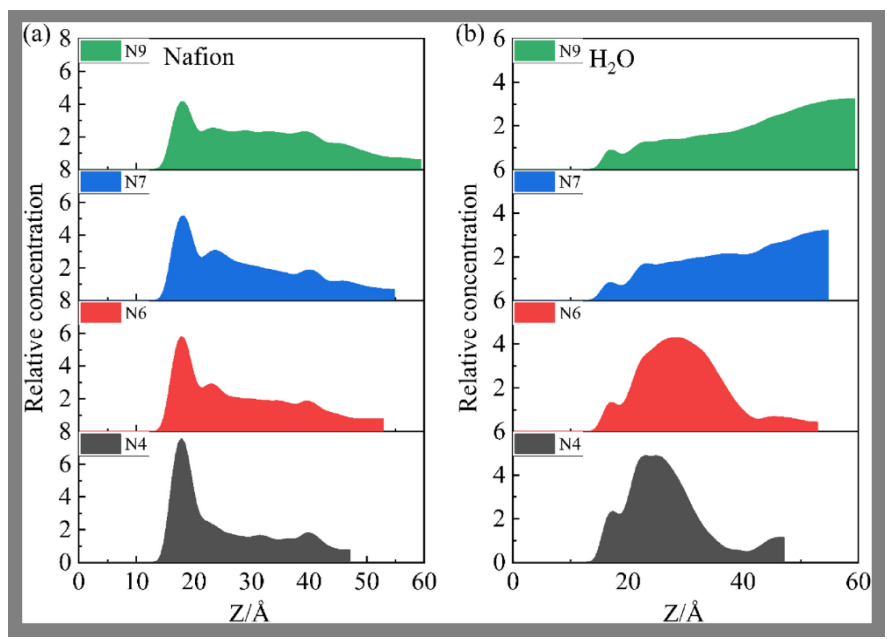

**Figure 16**: Relative concentration profiles in the Z-direction for each model with different Nafion numbers: (**A**) Nafion®; (**B**) $H_2O$. The N number indicates the number of Nafion polymers in the simulation structure. Reused with permission from.[170]

In studies on pure Pt surfaces, the Nafion layer shrinks due to strong adhesive forces pulling Nafion to the surface, limiting proton and water migration.[169] However, molecular hopping (where molecules move perpendicular to the surface) has a lower diffusion barrier. Such tight packing can be further enhanced by changes in humidity and temperature. The ionomer film also protects the CL from oxidation from $O_2$. This is because the Nafion controls and regulates the movement of $O_2$ and water, which limits its range of damage.[173] With a damaged ionomer layer, the water density grows close to the metal surface and increases the interfacial repelling force. This resistance stems from a "dense layer" formed near the Pt surface where sulfonic acid groups from the ionomer side chains adsorb strongly, creating a physical barrier to oxygen molecules. As the $SO_3$ density reduces, the barrier effect reduces. In all, it's a delicate balance between regulated flow of $O_2$/water and not overburdening the Pt surface.

The chemical bonds between the reactive anions (O⁻ and OH⁻) are dependent on the polymer structure.[174] If the sulfonate group adsorbs onto the Pt(111) surface, it generates two additional adsorbates ($OH_{naf_{ion}}$ and $O_{naf_{ion}}$) which are kinetically more sluggish. The new adsorbates hinder O* to OH* conversion and OH* reduction. The impact of a free $SO_3$ molecule would also presumably have a similar impact and limit the overall ORR kinetics. The cleavage of C–S bonds in the sulfonic acid functional group typically has low reaction barriers, which significantly impacts proton mobility and leads to mechanical failures like delamination. Understanding these degradation mechanisms is critical, as the breakdown of either Pt or Nafion initiates a feedback cycle that accelerates degradation at the interface.

Despite being a site of radical attack, DFT modeling also reveals that Nafion provides a protective effect for the Pt(111) surface against CO poisoning. Under the action of Nafion, the adsorption energy of CO on the Pt(111) catalyst increases by approximately 5 eV, which significantly enhances the catalyst's resistance to CO toxicity.[114] This resistance is attributed to the interaction between the hydroxyl (–OH) groups in the Nafion side chain and the CO molecules, which effectively reduces the strength of the interaction between CO and the Pt surface. Consequently, the stronger the adsorption performance of Nafion on the Pt(111) surface, the more robust the anti-CO poisoning capability of the catalyst becomes.

# 7. Nafion Alternatives and Emerging Membranes

A recurring theme throughout is that many of the most severe PEMFC degradation mechanisms are not incidental weaknesses of the proton exchange membrane but are intrinsic consequences of Nafion's hydration-dependent proton conduction mechanism. Nafion requires adequate hydration to function, narrowing the operating temperature window to below approximately 80 °C and necessitating complex water management systems that introduce additional failure modes.[175] This constraint motivates growing computational interests in alternative membrane materials capable of anhydrous or low-humidity product conduction. Critically, the same computational tools that are being utilized to illuminate Nafion degradation mechanisms are now being deployed to screen and characterize candidate replacement materials, making this an area where the computational advances discussed in this review translate directly into practical durability improvements.

## 7.1. Limitations of Nafion as a Degradation Prone Structure

Nafion's architecture is intrinsically vulnerable to the coupled degradation described throughout this review. Nafion's proton conductivity depends entirely on the formation of hydrophilic water channels through phase separation of its perfluorinated backbone and sulfonated side chains.[161]

Under dry conditions, these channels collapse, proton conductivity drops precipitously, and the membrane becomes mechanically brittle. Under high humidity, the membrane swells, generating the hygrothermal stresses that drive interfacial delamination and fatigue cracking. The sulfonated side chains that enable proton transport are simultaneously the primary sites of radical attack, meaning that the very functional groups responsible for membrane performance are those most vulnerable to degradation.[176,177] Furthermore, the hydration requirement constrains operating temperatures, preventing the kinetic benefits of higher-temperature operation and making cold-start behavior and freeze-thaw cycling unavoidable in automotive applications.

This architecture creates a fundamental tension: the conditions that maximize proton conductivity are close to those that maximize degradation rate. Computational modelling has clarified the molecular origins of this tension, but has also begun to identify materials where the proton conduction mechanism does not rely on hydration, and where this tension is therefore absent.

## 7.2. Anhydrous Proton Conduction

Among the most computationally promising candidates for anhydrous proton conduction are two-dimensional carbon frameworks functionalized with hydrogen-bonding groups. Hydroxylated graphane, or graphanol, was simulated, demonstrating anhydrous proton diffusion via Grotthuss-type hopping along a two-dimensional hydrogen-bonding network of hydroxyl groups.[178] The mechanism differs fundamentally from Nafion in that proton transport does not require the presence of liquid water. Alternatively, the ordered hydrogen-bond network of surface functional groups provides a percolating pathway for proton hopping at all hydration levels above the percolation threshold. Continuing this work Ananthabhotla, L. Y. et al. (2025) constructed a deep learning potential for graphamine using the DeepMD formalism combined with an active learning cycle based on the DP-GEN framework, achieving near DFT accuracy.[175] Using this deep learning potential, proton diffusion coefficients were computed across a temperature range of 300 – 500 K, yielding an estimated activation energy barrier of 63 ± 4 meV for anhydrous proton conduction. His value is substantially lower than that of graphanol (99 ± 13 meV) and significantly below the barriers reported for Nafion and other conventional membrane materials, including polybenzimidazole, polysulfone, and poly(ether ether ketone). The estimated proton conductivity of graphamine reaches 1322 mS/cm at 300 K, which exceeds the conductivity of Nafion under optimal hydrated conditions by more than an order of magnitude, and surpasses all other membrane materials of which the authors are aware.[175]

The mechanistic origin of graphamine's superior performance relative to graphanol involves the two-dimensional Grotthuss chains formed by hydrogen bonding between neighboring amine groups. The proton bearing $NH_3$ group can hop in three directions corresponding to three possible hydrogen bonds, compared with only two directions available on graphanol. This higher connectivity generates longer Grotthuss chains, more branched transport networks, and greater numbers of high-coordination nodes, collectively explaining why graphamine achieves lower apparent activation energies despite having larger intrinsic hopping and rotational barriers than graphanol.[175]

## 7.3. Broader Landscape of Computational Membrane Screening

Beyond two-dimensional carbon frameworks, a broader class of candidate materials is attracting computational attention. Metal-organic frameworks, covalent organic frameworks, and solid acid electrolytes have each been proposed as Nafion alternatives for high-temperature or anhydrous operation, and ML driven screening approaches are beginning to be applied to these material classes.[67] The common thread across these efforts is the use of graph neural network potentials and high throughput DFT to map structure property relationships across large chemical spaces, identifying candidates with favorable combinations of proton conductivity, chemical stability, and mechanical robustness before committing to expensive synthesis and characterization.

What remains largely absent from this literature is a comparative degradation study. A study that evaluates candidate materials not only on their conductivity under ideal conditions but also on their resistance to the coupled chemical, mechanical, and electrochemical degradation modes that ultimately determine PEMFC lifetime. This represents both the most important gap in the computational membrane screening literature and the most natural extension of the mechanistic degradation work reviewed in preceding sections. A unified computational framework that assesses both transport performance and degradation resistance would represent a qualitative advance in the predictive design of durable proton exchange membranes.

## 8. Conclusions and Future Direction

This review has examined PEMFC degradation through a computational lens, tracing the atomistic and molecular origins of chemical, electrochemical, mechanical, and contamination driven failure across multiple length and time scales. A central and recurring finding is that these degradation pathways do not operate independently. Instead, they form strongly coupled feedback loops in which each failure mode creates the conditions for the next. Radical attack cleaves sulfonated side chains in Nafion, simultaneously reducing proton conductivity and generating the carboxyl-terminated defects that lower the barrier for further backbone unzipping. Platinum dissolution under high potentials produces surface vacancies that act as preferential sites for hydrogen peroxide formation, renewing the radical species responsible for membrane chemical degradation. Carbon support corrosion lowers platinum detachment barriers, accelerating Ostwald ripening and exposing fresh carbon surface to further oxidative attack. Ionomer degradation at the catalyst membrane interface disrupts oxygen and water transport, altering local reaction conditions in ways that amplify both chemical and electrochemical degradation rates. Cationic and anionic contaminants simultaneously disrupt membrane proton transport, modify the electric double layer at the catalyst surface, and in the case of redox active species such as $Fe^{2+}$ and $Cu^{2+}$, directly catalyze the radical chemistry that initiates membrane scission.

What this review makes clear is that while computational modelling has achieved remarkable mechanistic resolution of individual degradation processes, the field has systematically studied these processes in isolation. No existing computational framework has been designed to capture this coupling simultaneously under realistic operating conditions. This gap between the mechanistic detail available for individual processes and the coupled reality of degradation in an operating fuel cell represents the defining unresolved challenge for the field, and it is the challenge that future computational work must be oriented toward.

Despite this limitation, the body of work reviewed here represents substantial progress. DFT has elucidated the energetics of platinum dissolution and oxide formation with atomic precision, identified the self-limiting nature of Pt dissolution under steady state conditions, and revealed the synergistic roles of radicals in Nafion backbone degradation. The development of constant-potential simulation approaches and the computational hydrogen electrode have begun to close the gap between idealized DFT calculations and electrochemically realistic boundary conditions, though significant challenges in describing the dynamic electrified interface remain. Molecular dynamics simulations have provided a mesoscale picture of Nafion morphology, water channel formation, ion transport, and the mechanical response of the membrane to hydration cycling and tensile stress, with modern force fields achieving quantitative agreement with experimental measurements of elastic modulus and proton diffusivity. Reactive force fields, despite their known transferability limitations, have enabled the first explicit simulations of bond-breaking degradation events in Nafion and at the platinum carbon interface. Machine learning potentials have demonstrated their capability to describe Pt nanoparticle dissolution, oxidation state evolution, and in the case of graphamine anhydrous proton conduction, with near DFT accuracy at computational costs compatible with the system sizes and timescales relevant to degradation.

The contaminant literature has benefited substantially from this computational progress. The mechanistic classification of cationic contaminants into electrostatic crosslinking, hydration-dominated, redox driven, and electrochemical modulation regimes as developed in this review provides a framework for understanding why species as structurally similar as $Ca^{2+}$ and $Mg^{2+}$ produce mechanistically distinct degradation responses, and why redox-active species such as Fe and Cu dominate long-term degradation despite their lower environmental prevalence. For anionic and gaseous contaminants, DFT has clarified the surface chemistry underlying chloride accelerated Pt dissolution, sulfur poisoning of ORR active sites, and the CO tolerance mechanism conferred by intact Nafion at the Pt(111) surface.

Despite this progress, several gaps stand out as both practically significant and tractable with near term computational advances. The coupled degradation of the Nafion Pt carbon triple-phase boundary where chemical, mechanical, and electrochemical processes converge has not been modelled within any single computational framework. The degradation transport feedback, in which structural damage to the ionomer alters oxygen and water mobility in ways that further accelerate degradation, remains entirely uncharacterized at the molecular level. The electrochemical potential dependence of radical generation, Pt dissolution, and ionomer degradation has not been integrated into a single simulation, meaning that the potential cycling conditions most relevant to real automotive and stationary applications are still beyond the reach of existing methods. The behavior of realistic, defects containing membrane and catalyst structures is poorly understood, and the role of manufacturing variability in determining degradation onset has received almost no computational attention.

The contaminant literature, while extensive experimentally, remains mechanistically fragmented in simulation. Few studies have systematically compared binding free energies, diffusion coefficients, hydrogen-bond network perturbations, or electric double layer modifications across different contaminant species within a consistent membrane model. Without such comparative frameworks, it is not currently possible to rank contaminant severity from first principles or to predict synergistic effects when multiple contaminants are present simultaneously, as is typically the case in real operating environments.

The path toward predictive, coupled degradation modelling runs through machine learning interatomic potentials. The development of next generation MLIP architectures has produced potentials with the accuracy, transferability, and computational efficiency required to simulate the chemically heterogeneous PEMFC interface at relevant length and time scales. Several principles should guide their development for PEMFC applications.

Training data quality and diversity is the single most important determinant of MLIP performance and deserves careful attention from the outset. Training sets must sample not only equilibrium configurations but transition states, partially degraded structures, and rare reactive events including radical attack and bond scission. Static DFT datasets are insufficient for this purpose because they systematically under sample the reactive configurations most relevant to degradation. Active learning workflows are the most efficient strategy for building transferable datasets, as demonstrated for graphamine and carbon nanostructure growth on metallic surfaces. Researchers entering this field should implement active learning from the beginning of potential development rather than treating it as a refinement step.

The explicit treatment of electrochemical potential within MLIP frameworks remains the most significant technical barrier to realistic degradation simulation. The majority of existing potentials are trained at fixed, neutral charge states and cannot capture the potential-dependent behavior governing Pt dissolution, oxide formation, and radical generation under operating conditions. Recent progress in constant-potential MD and charge-equilibration schemes offers routes toward electrochemically aware potentials, and their integration with equivariant architectures should be a priority for the field. The development of community curated training datasets for PEMFC relevant interfaces would accelerate this progress by reducing duplicated effort and enabling systematic benchmarking.

The emerging computational interest in Nafion alternatives, exemplified by deep learning potential studies of graphamine and graphanol, points toward an important broadening of the field's scope. Materials that conduct protons anhydrously through extended Grotthuss chains in two-dimensional hydrogen-bonding networks circumvent the hydration dependent degradation modes that dominate Nafion failure. The predicted proton conductivity of graphamine exceeds that of Nafion by more than an order of magnitude under optimal conditions, and its activation barrier of 63 meV is substantially below that of all conventional membrane materials studied to date. Whether these properties survive experimental synthesis, realistic defect concentrations, and the chemical and mechanical stresses of PEMFC operation remains to be established, and computational characterization of these materials under degradation relevant conditions is an urgent priority.

Finally, multiscale integration remains the long-term architectural goal for the field. The challenge is not merely computational but conceptual: developing physically consistent ways to pass information between scales without losing the mechanistic resolution that makes atomistic simulation valuable in the first place. Large in situ characterization datasets, including those now becoming available for cathode catalyst degradation under accelerated stress testing, provide the experimental benchmarks against which these frameworks must be validated. As the quality and diversity of both computational and experimental datasets grows, the field moves closer to the goal that motivates all of this work: predictive models of PEMFC lifetime that can guide the design of more durable membranes, catalysts, and membrane electrode assemblies for the stationary and automotive applications on which the hydrogen economy depends.

# Competing interests

The authors declare no competing interests in the writing of this review.

# Acknowledgements

The facilities at Kansai university, were imperative for the resources and access to literature to make this review. The authors would like to thank the JSPS organisation for their support with the grant "Japan Society for the Promotion of Science" (Number: 25KF0283), and fellowship award (Number: P25711).

Received: ((will be filled in by the editorial staff))
Revised: ((will be filled in by the editorial staff))
Published online: ((will be filled in by the editorial staff))